\documentclass[sigconf]{acmart}

\usepackage{makecell}
\AtBeginDocument{%
  }

\copyrightyear{2025}
\acmYear{2025}
\setcopyright{acmlicensed}\acmConference[CHI '25]{CHI Conference on Human Factors in Computing Systems}{April 26-May 1, 2025}{Yokohama, Japan}
\acmBooktitle{CHI Conference on Human Factors in Computing Systems (CHI '25), April 26-May 1, 2025, Yokohama, Japan}
\acmDOI{10.1145/3706598.3714277}
\acmISBN{979-8-4007-1394-1/25/04}

\acmConference[CHI '25]{The ACM CHI Conference on Human Factors in Computing Systems}{April 26--May 1, 2025}{Yokohama, Japan}

\begin{document}

\title[RemiHaven]{RemiHaven: Integrating ``In-Town'' and ``Out-of-Town'' Peers to Provide Personalized \textcolor{black}{Reminiscence Support} for Older Drifters}

\author{Xuechen Zhang}
\email{23210240386@m.fudan.edu.cn}
\affiliation{%
  \institution{Fudan University}
  \city{Shanghai}
  \country{China}
}

\author{Changyang He}
\email{changyang.he.young@gmail.comn}
\affiliation{%
  \institution{Max Planck Institute for Security and Privacy}
  \city{Bochum}
  \country{Germany}
}

\author{Peng Zhang}
\authornote{Corresponding authors.}
\email{zhangpeng_@fudan.edu.cn}
\affiliation{%
  \institution{Fudan University}
  \city{Shanghai}
  \country{China}
}

\author{Hansu Gu}
\email{guhansu@gmail.com}
\affiliation{%
    \institution{Independent}
  \city{Seattle}
  \country{USA}
}

\author{Ning Gu}
\email{ninggu@fudan.edu.cn}
\affiliation{%
  \institution{Fudan University}
  \city{Shanghai}
  \country{China}
}

\author{Qi Shen}
\email{qishen@fudan.edu.cn}
\affiliation{%
  \institution{Fudan University}
  \city{Shanghai}
  \country{China}
}

\author{Zhan Hu}
\email{zhanhu@fudan.edu.cn}
\affiliation{%
  \institution{Fudan University}
  \city{Shanghai}
  \country{China}
}

\author{Tun Lu}
\authornotemark[1]
\email{lutun@fudan.edu.cn}
\affiliation{%
  \institution{Fudan University}
   \city{Shanghai}
   \country{China}
}

\renewcommand{\shortauthors}{Xuechen Zhang et al.}

\begin{abstract}

With increasing social mobility and an aging society, more older adults in China are migrating to new cities, known as ``older drifters''. Due to fewer social connections and cultural adaptation, they face negative emotions such as loneliness and depression. While reminiscence-based interventions have been used to improve older adults' psychological well-being, challenges such as the lack of tangible materials and limited social resources constrain the feasibility of traditional reminiscence approaches for older drifters. To address this challenge, we designed RemiHaven, a personalized \textcolor{black}{reminiscence support tool} based on \textcolor{black}{a two-phase formative study.} It integrates ``In-Town'' and ``Out-of-Town'' peer agents to enhance personalization, engagement, and emotional resonance in the reminiscence process powered by Multimodal Large Language Models (MLLMs). Our evaluations show RemiHaven's strengths in supporting reminiscence while identifying potential challenges. We conclude by offering insights for the future design of \textcolor{black}{reminiscence support tools} for older migrants.

\end{abstract}



\maketitle

\section{INTRODUCTION}
With the acceleration of social mobility and an aging society, an increasing number of older adults are migrating to new cities, particularly in developing countries such as China \cite{liu2019nothing, tang2022social}, the Philippines \cite{montayre2017moving}, and Mexico \cite{acharyavulnerability}. This migration is often driven by the desire to reunite with family or relieve the burden of their children \cite{liu2019nothing, tang2022social}, leading to the rise of a special elderly group known as the ``older drifters'' (``lao piao'' in Chinese). These are defined as older adults who leave their hometowns later in life to live in cities where their children reside \cite{lei2024unpacking}. According to the Chinese Migration Report, over the past 20 years, the number of late-life migrants in China has increased from 5 million to 18 million, among whom 77\% are older drifters \cite{unicef2019china, xinhua2018migrant}. Compared to the \textcolor{black}{older migrants who have received more attention in the HCI field \cite{liaqat2022hint}}, older drifters face specific challenges such as fewer social connections, lower digital literacy, and greater difficulty adapting to new cultures \cite{da2015later, tang2022social}. These obstacles make it difficult for them to engage in unfamiliar urban environments, often leading to psychological challenges like loneliness, a sense of loss, and depression \cite{liu2016dilemma, tang2022social, wang2022mental}. 

Reminiscence-based intervention is a common psychosocial intervention approach to improve the well-being of older adults \cite{tam2021effectiveness}. It can release depressive emotions, increase satisfaction, and promote self-esteem \cite{tam2021effectiveness}. Additionally, appropriate reminiscence enhances cognitive functions like memory and concentration \cite{cotelli2012reminiscence}, contributing to healthy aging \cite{butler1963life}. \textcolor{black}{For older drifters, reminiscence is also very important, since it can help maintain self-identity and emotional connections as well as alleviate the psychological discomfort caused by migration. Research suggests that environmental memories could bridge the past and present, thus fostering psychological integration \cite{chaudhury1999self}. Recalling familiar scenes and places could help older drifters who live in unfamiliar environments and experience a loss of social networks find a sense of belonging and security \cite{jooj2016effect}.} For reminiscence-based intervention, the most widely adopted method is reminiscence conversation, which uses tangible materials such as old photographs or heirlooms to evoke memories and stimulate conversation \cite{bulechek2013nursing}. These conversations can be conducted in groups or individually, eventually forming some forms of life storybook \cite{woods2018reminiscence}. However, migration often results in a lack of both tangible reminiscence materials and access to social work resources, further limiting the effectiveness of these intervention methods within this group. This problem inspires us to explore: \textcolor{black}{\textcolor{black}{How can} we design a reminiscence support tool with personalized experience to address older drifters' lack of tangible materials and other potential barriers in reminiscence?}

The development of Large Language Models (LLMs), especially multimodal models (MLLMs), presents an opportunity to address the aforementioned question. Recent research has explored the feasibility of using generative AI to support music-based reminiscence for older adults \cite{jin2024exploring}, showing that generative AI can enhance reminiscence by generating conversations and images. Building on this, rather than simply leveraging the interactive and content-generating capabilities of LLMs, we focus on customizing MLLMs to better address the specific reminiscence needs of older drifters. By tailoring MLLMs to operate effectively when tangible materials are scarce or unavailable, we aim to generate personalized and contextually relevant multimodal content, including conversations, images, and text that resonates with the life experiences of older drifters, creating a more effective and engaging reminiscence process.

For this study, we conducted three main works. First, a two-phase formative study was conducted to systematically understand older drifters' current reminiscence practices and their expectations for future designs. In the first phase, semi-structured interviews revealed unique challenges faced by older drifters in their reminiscence practices, including a lack of tangible reminiscence materials and the difficulty of sharing memories with others due to disrupted social networks. The second phase, through storyboard-based user studies, further explored participants’ expectations for a reminiscence support tool. This iterative approach identified five key design goals, such as allowing users to configure the level of control for the reminiscence assistant.

Second, based on these design goals, we developed and refined RemiHaven, \textcolor{black}{a reminiscence support tool} specifically designed for older drifters. This tool centers around a conversational agent powered by an MLLM, integrating two core modules: the Prompt Organization module, which aggregates and structures contextual information from conversations and background details, and the Memory-Material Generation module, which creates personalized multimodal content, including images and text. A key innovation of RemiHaven is its support for two distinct types of interaction modes: Conversation with ``In-Town'' Peers and Conversation with ``Out-of-Town'' Peers. These character-based interactions enrich the emotional resonance and overall reminiscence experience by simulating both familiar and novel social settings. The generated content can be compiled into a personalized life storybook, which can be easily exported and shared, offering users an engaging and accessible way to preserve cherished memories \cite{lindley2012before}.

Finally, we recruited 10 older drifters to evaluate RemiHaven's effectiveness in \textcolor{black}{supporting reminiscence, employing both questionnaires and semi-structured interviews with two sets of metrics: the Technology Acceptance Model (TAM) \cite{marangunic2015technology} and the system's alignment with design strategies.} Results from both quantitative and qualitative data showed that RemiHaven significantly improves the reminiscence experiences and \textcolor{black}{contributes to participants' positive emotional responses}. These findings revealed key insights for designing \textcolor{black}{reminiscence support tools} for older drifters. Furthermore, the broader implications of generative AI in reminiscence \textcolor{black}{were discussed.}

In conclusion, our contributions can be summarized as follows:
\begin{itemize}
    \item To the best of our knowledge, this is the first study to explore a \textcolor{black}{reminiscence support tool} for older drifters.
    \item We conduct a formative study and uncover older drifters' special requirements for designing such a tool.
    \item We design and implement \textcolor{black}{a reminiscence support tool} for older drifters based on MLLMs, and validate its effectiveness in enhancing reminiscence experiences and \textcolor{black}{contributing to emotional benefits} among participants.
    \item We propose several insights into the future research and design of such systems.
\end{itemize}

\section{RELATED WORK}
\label{Related Works}

\subsection{Challenges and Opportunities in Supporting Late-Life Migrants}
Older migrants include individuals who migrate at an older age and younger adults who have aged in their destinations \cite{warnes2006older}, including both domestic migrants \cite{unicef2019china, xinhua2018migrant} and international migrants \cite{maleku2022we, du2023understanding}. Late-life migrants, a specific subset of older migrants, are those who relocate later in life \cite{warnes2006older}. Their motivations vary, including family reunification \cite{maleku2022we, montayre2017moving}, seeking better living environments \cite{montayre2017moving, du2023understanding}, caregiving for children \cite{da2015later, zhou2012space}, and responding to social changes \cite{heikkinen2013transnational}. With increasing urbanization and aging populations, late-life migration has emerged as a prominent global phenomenon \cite{migrationdata2023olderpersons}.

In China, influenced by traditional family culture \cite{sun2010value}, an increasing number of older adults are migrating to large cities to support their children or to reunite with families \cite{liu2019nothing, tang2022social}. These late-life migrants known as ``older drifters'' (``lao piao'' in Chinese), are becoming more prevalent with the increasing social mobility of the younger generation. The latest report shows that, over the past 20 years, the population of older migrants in China has increased from 5 million to 18 million, with 77\% classified as older drifters \cite{unicef2019china, xinhua2018migrant}. Due to fewer social connections, lower digital literacy, and cultural adaptation challenges \cite{da2015later, tang2022social}, older drifters often struggle to integrate into urban life, facing psychological challenges such as loneliness, a sense of loss, and depression \cite{liu2016dilemma, tang2022social, wang2022mental}. These issues have attracted attention from sociology and gerontology, particularly regarding their social adaptation, community participation, mental health, and quality of life \cite{bao2022embedding, liu2023association, tang2022social}.

In HCI field, although aging and migration have received increasing attention, research on older migrants, particularly older drifters, remains limited \cite{lei2024unpacking}. Tang and Chandra \cite{tang2022community} studied the economic practices of older migrants in Hong Kong, revealing the importance of social capital and community networks in helping their adaptation to new environments. Liaqat et al. \cite{liaqat2022hint} explored how storytelling in multigenerational migrant families facilitates intergenerational communication between older migrants and their grandchildren. Building on these studies, Lei et al. \cite{lei2024unpacking} introduced the concept of ``older drifters'' into HCI through a qualitative study, using the social convoy model to reveal the challenges such as cultural adaptation and weakened social connection. They also proposed several Information and Communication Technology (ICT) design considerations for this group. However, despite these efforts, there remains a lack of ICT design solutions specifically tailored to the needs of older drifters.

Our work aims to explore how HCI design can address the unique mental health challenges faced by older drifters. It seeks to fill the gaps in existing research by providing support tools for this rapidly growing migrant group, \textcolor{black}{specifically through the lens of reminiscence, which contributes to their emotional benefits with fewer constraints and social resources.}

\subsection{Improve the Well-being of the Elderly through Reminiscence-based Intervention}
\label{2.2}
Reminiscence-based intervention is a common psychosocial intervention to enhance positive emotions, alleviate depression, and improve self-esteem and life satisfaction among older adults by recalling, narrating, and reflecting on past experiences \cite{woods2018reminiscence}. This process typically involves tangible materials, such as old photographs or heirlooms, to evoke memories and stimulate conversations \cite{mccloskey2000nursing}. By sharing experiences and creating some forms of life storybook \cite{woods2018reminiscence}, reminiscence-based intervention can also improve cognitive function and personal identity, further enhancing the life quality of older adults \cite{huang2015reminiscence}. These interventions can be conducted individually or in groups across various settings \cite{gaggioli2014effectiveness}, making them effective, low-risk, and easy to implement \cite{jooj2016effect}. While reminiscence-based intervention shares similarities with reminiscence therapy \cite{bartels2004evidence, song2014effects}, these two concepts are sometimes equivalent in research \cite{tam2021effectiveness}. Nevertheless, reminiscence-based intervention is typically used for older adults without clinical mental disorders, serving as a preventative measure for mental health issues.

In HCI, reminiscence-based intervention support techniques have become an important research direction, particularly in the design of tools for older adults. Researchers and designers have developed various tools to aid memory recall, such as multi-touch tablets and photo viewers, which support sharing memories between older adults and their families \cite{apted2006tabletop, kirk2010opening, axtell2019photoflow}; immersive VR techniques to facilitate collaborative reminiscence activities \cite{baker2021school, shen2024legacysphere}; and sensory stimuli via sounds and music to enhance the reminiscence process \cite{jin2024exploring, odom2019investigating}. However, most research and design efforts have primarily focused on older adults with some cognitive impairments, centering more on reminiscence therapy.

While these studies provided various methods and tools to support reminiscence for older adults, their applicability to older drifters remains uncertain. \textcolor{black}{For this group, memories about familiar places play a crucial role in anchoring self-identity, fostering a sense of belonging and psychological security \cite{chaudhury1999self}. Reminiscence can help older drifters mitigate the psychological challenges of migration by aiding adaptation to unfamiliar environments and maintaining identity continuity \cite{LiuQing2012OldDriftingGroup}. However, their lack of tangible materials and limited social resources might hinder the applicability of existing systems.} Therefore, designing suitable \textcolor{black}{reminiscence support tools} for this group has become an urgent issue.

\subsection{Multimodal Large Language Models for HCI}
In recent years, MLLMs have gained increasing attention in various fields \cite{yin2023survey}. These models are not only capable of understanding and processing complex multimodal data, such as text, images, and audio, but are also effective at generating content consistent with human expectations. Tools like ChatGPT, DALL·E, and Midjourney have expanded possibilities across HCI domains, including education, design, and eldercare. For instance, in education, Suresh et al.'s ``StoryBuddy'' enhances children's language skills by generating tailored parent-child storytelling content \cite{zhang2022storybuddy}. In design, Lin and Martelaro's ``Jigsaw'' system leverages the capabilities of MLLMs to simplify task-switching to streamline the creative process \cite{lin2024jigsaw}. In eldercare, Xygkou et al. designed ``MindTalker'', a GPT-4-powered conversational agent to support social interaction for dementia patients \cite{xygkou2024mindtalker}. These examples highlight MLLMs' growing role in enhancing human interaction and creativity.

For older drifters, who face limited access to tangible reminiscence materials and require personalized and engaging experiences, MLLMs offer a promising solution. MLLMs can capture and respond to the reminiscence needs by asking relevant questions, guiding detailed recall, and generating personalized content, such as text and images, to support reminiscence.

To our knowledge, few studies have applied MLLM-related techniques to support reminiscence. One recent study explored generative AI for music-based reminiscence among older adults \cite{jin2024exploring}, examining their attitudes towards using AI to support reminiscence and providing valuable insights into future design. However, this research focused on \textcolor{black}{ investigating the use of music as a reminiscence trigger for older adults}. By contrast, our study focuses on older drifters, exploring how MLLMs can support reminiscence for this group. We aim to build a user-centered tool that enriches and personalizes their reminiscence experiences, uncovering new perspectives and possibilities for alleviating older drifters' emotional problems in the new environment.

\section{FORMATIVE STUDY}
\label{Formative Study}
As mentioned in Section~\ref{2.2}, while existing studies have made significant progress in building tools that support reminiscence for older adults, these solutions may not sufficiently address the specific needs of older drifters. \textcolor{black}{To bridge this gap, we conducted a two-phase formative study focused on older drifters' current reminiscence practices and expectations, respectively.} \textcolor{black}{The first phase utilized semi-structured interviews to investigate the reminiscence characteristics and current tool usage patterns of older drifters, uncovering their unique challenges. Building on these findings, the second phase employed storyboard-based user studies to dive into their expectations for \textcolor{black}{reminiscence support tools}.}

\subsection{Method}
\subsubsection{Participants}

We recruited older drifters through announcements and outreach in three diverse communities in Shanghai to ensure a representative sample. Participants met the inclusion criteria of being 60 years or older and having relocated to Shanghai with their children for over 3 months. \textcolor{black}{We did not require participants to have prior experience with reminiscence support tools, as this study aimed to explore general reminiscence needs and experiences of older drifters, rather than focusing on those with advanced digital literacy or familiarity with ICT tools.} Ultimately, we obtained 13 older drifters (8 women and 5 men, aged 62 to 78). The gender distribution aligns with prior work \cite{yue2017}, which suggests that women are more likely to relocate compared to men, often to assist in caring for grandchildren. Participant details are presented in Table \ref{tab: Summary of participants}.

\textcolor{black}{For the second phase, we invited all participants from the first phase to join the storyboard-based user studies.} Except for P3, P4, and P7, who had scheduling conflicts, the remaining 10 participants agreed and participated. All expressed a need for reminiscence and showed strong interest in the reminiscence support tool we aimed to design.

\begin{table}[ht] \small
\caption{Detailed information of participants.}
\label{tab: Summary of participants}
\centering
\begin{tabular}{ccccccc}
\hline
\textbf{ID} & \textbf{Gender} & \textbf{Age} & \textbf{Hometown} & \textbf{Duration Since Relocation}\\ 
\hline
P1  & Female & 66-70 & Heilongjiang & 9 years \\
P2  & Female & 61-65 & Jilin        & 10 years \\
P3  & Male   & 71-75 & Jiangsu      & 15 years \\
P4  & Female & 66-70 & Jiangsu      & 4 months \\
P5  & Female & 66-70 & Shanxi       & 2 years \\
P6  & Female & 66-70 & Jiangxi      & 16 years \\
P7  & Male   & 71-75 & Shandong     & 6 years \\
P8  & Male   & 71-75 & Jiangxi      & 16 years \\
P9  & Female & 61-65 & Jiangsu      & 17 years \\
P10 & Female & 66-70 & Shanxi       & 20 years \\
P11 & Male   & 76-80 & Shandong     & 25 years \\
P12 & Female & 66-70 & Shaanxi      & 8 years \\
P13 & Male   & 61-65 & Anhui        & 12 years \\
\hline
\end{tabular}
\end{table}

\subsubsection{Procedure}
\begin{figure}[ht]
    \centering
    \includegraphics[width=0.8\linewidth]{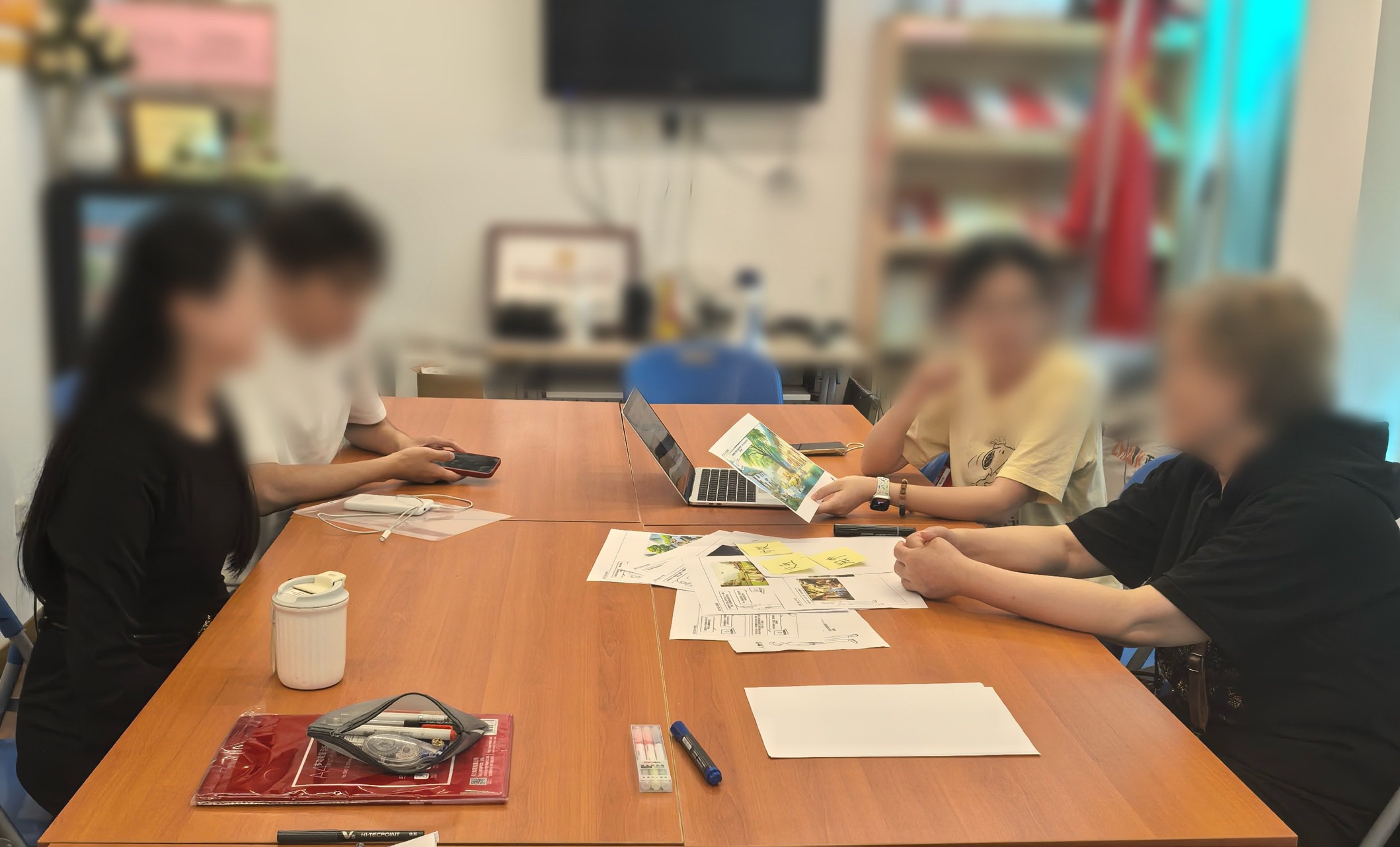}
    \caption{The activities conducted during storyboard-based user studies.}
    \Description{Participants engaging in user study activities focusing on reminiscence support tool design. Four individuals are seated around a table, reviewing storyboards that depict different reminiscence scenarios and discussing potential improvements. Various documents and materials, including a laptop and printed storyboards, are spread across the table. The session emphasizes gathering user feedback and refining design concepts based on the participants’ preferences and suggestions.}
    \label{fig: PD}
\end{figure}

\begin{figure*}[ht]
    \centering
    \includegraphics[width=\linewidth]{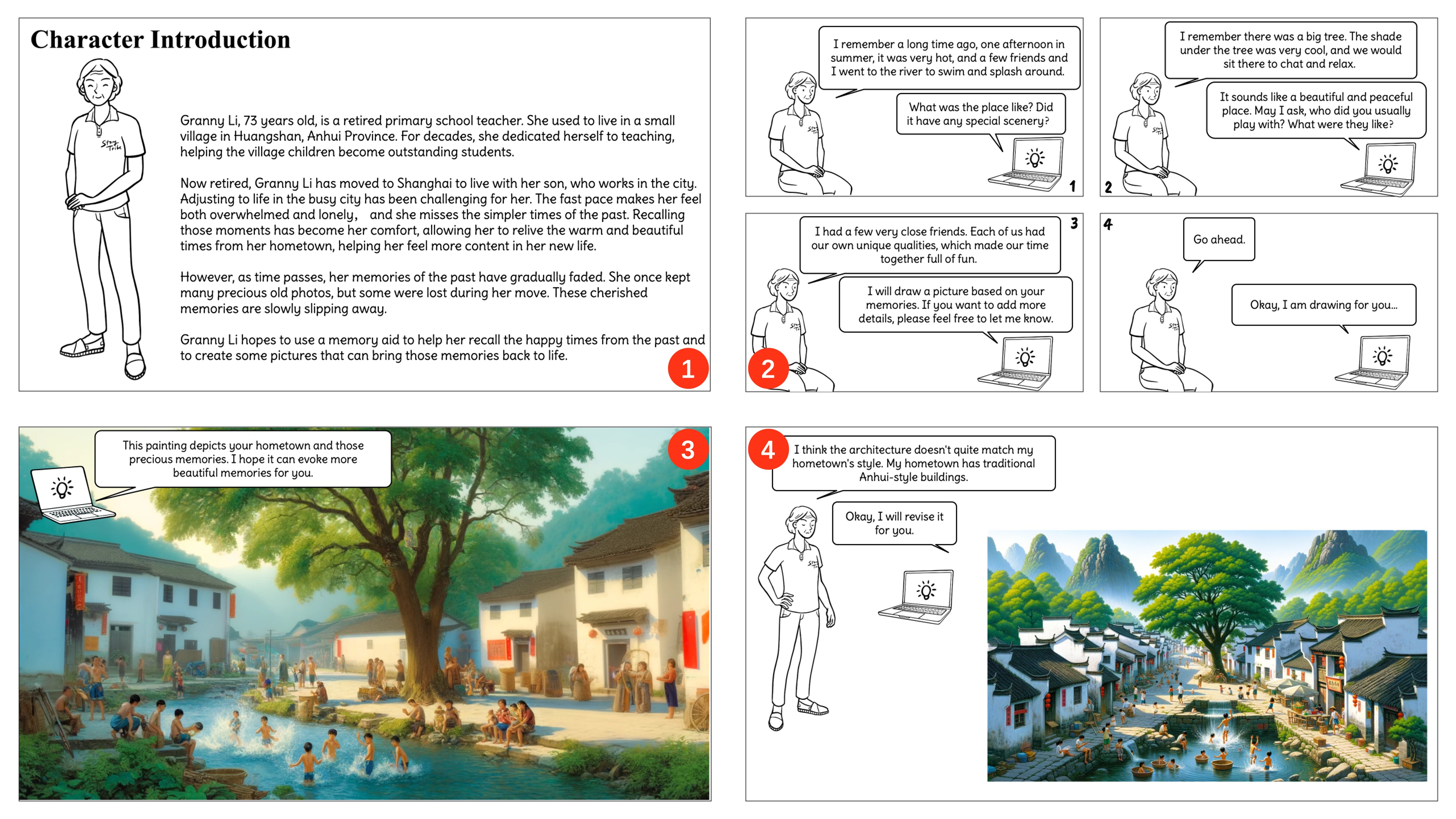}
    \caption{An example storyboard illustrating reminiscence scenarios tailored for older drifters.}
    \Description{This example storyboard presents reminiscence scenarios tailored to older drifters, focusing on their interactions with a reminiscence assistant. The storyboard highlights features such as user configuration based on demographic inputs, the choice between interaction modes like `In-Town'' and `Out-of-Town'' Peers, and multimodal content generation using MLLMs. It illustrates the assistant's ability to generate personalized prompts and memory materials, including images and text, and demonstrates how users can refine generated content and compile it into a life storybook.}
    \label{fig: storyboard}
\end{figure*}

\textcolor{black}{In the first phase,} we conducted semi-structured interviews to explore the reminiscence characteristics of older drifters, including the frequency, content, and methods. Interviews also explored their usage and attitudes toward existing reminiscence methods and tools. Each interview began with open questions with fewer restriction, encouraging free discussion of reminiscence-related experiences, \textcolor{black}{such as the role of reminiscence in their daily lives, the individuals with whom they usually share their memories, and any challenges they encountered in reminiscing or using existing tools.} All interviews were conducted in community centers and lasted 30 minutes.
\textcolor{black}{Based on the findings of the first phase (as described in Section \ref{findings}), the second phase employed storyboard-based user studies to delve into participants’ expectations for reminiscence support tools. These storyboards, designed based on personas derived from the first phase findings, prioritized the interactions between users and the reminiscence assistant rather than interface details \cite{gruen2000storyboarding, grudin2002personas, zhang2022storybuddy}. This approach was inspired by participatory design practices \cite{vines2012cheque, zhao2024older, liaqat2022hint}, emphasizing user feedback and preferences to refine design concepts.} During this phase, as shown in Figure \ref{fig: PD}, participants were presented with three low-fidelity storyboards depicting different reminiscence scenarios for older drifters. \textcolor{black}{The storyboards (Figure \ref{fig: storyboard} shows translated examples) focused on two key aspects: interaction features, such as the initiative and modalities of a reminiscence assistant and the presentation of reminiscence content, including image generation, modifiability, and the creation of a life storybook.} We conducted these sessions in community centers or participants’ homes, each lasting 30 to 45 minutes. \textcolor{black}{Participants were encouraged to discuss their feelings about each scenario, share preferences, and propose enhancements.} After presenting the storyboards, if participants were interested, we also assisted them in trying out MLLMs, including state-of-the-art models: ChatGPT-4 for dialogue generation and DALL·E 3 for image generation. Each participant received 50 yuan as compensation.


With the participants’ consent, interviews were audio-recorded and transcribed using automated tools and then verified by the first author. \textcolor{black}{Both phases followed a thematic analysis process to extract insights from the collected data \cite{khokhar2020theory, belotto2018data, belur2021interrater}. Supported by MAXQDA \footnote{MAXQDA: https://www.maxqda.com}, a qualitative data analysis software, the analysis involved systematically organizing, coding, and comparing data segments. Initially, two authors independently reviewed all feedback to familiarize themselves with the data and identify key concepts. Three rounds of iterative coding were conducted to refine the coding scheme. The first round focused on initial open coding, capturing key concepts and patterns that naturally emerge from the data. In the second round, initial codes with similar meanings were grouped into broader themes, with weak or redundant themes removed. The third round involved collaboratively refining and finalizing themes, ensuring coherence and alignment with the research objectives. Once the codebook was finalized, a sociology expert reviewed the codes to confirm thematic coverage, evaluate the logical structure, and suggest improvements. To ensure reliability, 30\% of the transcripts were randomly selected and independently coded by two coders, using the established codebook without prior joint discussion. The discriminant capability of the coding scheme was evaluated using Campbell et al.'s method \cite{campbell2013coding}. The inter-rater reliability for this sample was 89.8\%, \textcolor{black}{which falls within the commonly accepted threshold of 85\% to 90\% \cite{miles1984qualitative}, indicating a strong level of agreement.} The primary insights discussed in Section~\ref{findings} directly stem from this thematic analysis, with each insight supported by well-defined coding themes.}

\subsubsection{Ethical Consideration}
\label{ethical}
Before each interview, we explained the study’s purpose, procedures, potential risks, and participants' rights in detail to ensure full understanding. Participants then signed informed consent forms, agreeing to participate and allowing data recording. Given the vulnerability of older adults, interview questions were reviewed by community workers, and a gerontology expert accompanied the interviews to ensure the reminiscence content did not negatively impact participants. To protect privacy, all data will remain confidential and not be released.

\subsection{\textcolor{black}{Findings: Key Themes}}
\label{findings}

\subsubsection{\textcolor{black}{Findings of the First Phase: Exploring Reminiscence Characteristics and Existing Tool Usage}}
\
\newline
\textbf{\textcolor{black}{F1-1\footnote{\textcolor{black}{We use "F" to denote findings from our formative study, where "F1" corresponds to Phase I and "F2" corresponds to Phase II.}}: The reminiscence of older drifters, especially when involving memories of their hometowns, plays a crucial role in their psychological well-being.}} Most of the 13 participants frequently engaged in reminiscence, with only 2 stating they rarely did. Our interviews suggest that recalling the past provided psychological meaning, offering both guidance for the future and emotional comfort. For instance, P10 mentioned that despite her past hardships, those experiences were valuable and helped guide her future decisions: \textcolor{black}{\emph{``The paths we walk always leave traces, and these experiences can help with future life.''} Additionally, participants frequently reflected on memories of their hometowns and emphasized how such reminiscence maintained their cultural roots and affirmed their sense of self, especially after relocation.} As P1 said, \emph{``After retiring and moving to Shanghai, I couldn’t live as carefree as I used to, so I often reminisce about the days back in my hometown.''} \textcolor{black}{These reflections can not only alleviate homesickness, but also help to connect past experiences to present lives and reinforce participants' identities \cite{chaudhury1999self}. P9 remarked, \emph{``When I couldn't go back during the pandemic, thinking about my hometown made me feel less sad. It reminded me of who I am and where I come from.''}}

\textbf{\textcolor{black}{F1-2: Migration generally results in the loss of reminiscence materials and disruption of social networks, hindering older drifters’ reminiscence practices.}} \textcolor{black}{Through the interviews, we observed that most participants engaged in relatively simple reminiscence practices. One major reason is that older adults' reminiscence practices usually depend on tangible triggers, such as photos or memorabilia, while these materials are generally inaccessible to older drifters due to migration.} As P9 explained, \emph{``I don’t have any old photos left. They were all lost.''} \textcolor{black}{And the difficulty of revisiting old places and rapid urban development in their hometowns further erased familiar landmarks, reducing the relevance of traditional reminiscence methods. As P13 noted, \emph{``The old street where we lived has been redeveloped. The old places are gone.''}} \textcolor{black}{Besides, migration can lead to disruption of older drifters' existing social networks and make it challenging for them to rebuild connections in new environments, which can also hinder older drifters' reminiscence practices. P7 highlighted this difficulty, noting that sharing with those unfamiliar with or uninterested in their hometown \emph{``feels meaningless because they can’t truly understand or appreciate it.''}}

\textbf{\textcolor{black}{F1-3: Older drifters are generally characterized by low digital literacy and prefer interactive and intuitive reminiscence approaches.}} \textcolor{black}{Consistent with previous studies \cite{da2015later, tang2022social}, we found that many participants had low digital literacy and confidence in using ICT tools, and they usually do not have enough time (e.g., need to take caregiving) to explore such tools. As P1 explained, \emph{``I’m too busy taking care of my grandchildren. I don’t have the energy to learn how to use complex software.''} Due to these challenges, participants expressed a strong desire for interactive and intuitive reminiscence methods, which are easy to use.} Seven participants favored interactive systems that simulate conversational sharing, as P5 noted, \emph{``What would be practical is a way to chat and interact during reminiscence.''} Six participants preferred intuitive approaches, such as videos, images, or VR, to stimulate memories and make sharing easier with younger generations. As P10 mentioned, \emph{``My grandchildren only like to watch videos and pictures. I think this way is more intuitive and better than text.''} 

\subsubsection{\textcolor{black}{Findings of the Second Phase: Exploring Expectations for Reminiscence Support Tools}}
\
\newline
\textcolor{black}{\textbf{F2-1: Older drifters expect a reminiscence assistant that can play two different roles.} Through storyboard discussion, we found that participants envisioned a reminiscence assistant playing two primary roles: a guider who is familiar with their hometowns and a listener who is interested in their hometowns. Six participants preferred an assistant who is like a familiar person or someone from their hometowns.} Such an assistant would understand their hometown background and take a more active role in guiding the conversation, helping them organize their thoughts with questions and clues. As P5 said, \emph{``I think it would be better to have some direction because I tend to ramble.''} Similarly, P12 found it hard to articulate scattered memories clearly without such guidance. Conversely, some more confident participants preferred a freer conversational style, where they could share their memory independently. \textcolor{black}{In this scenario, the assistant should express interest in their memories and listen attentively, encouraging them to delve into their recollections at their own pace.} \textcolor{black}{As P6 remarked, \emph{``When someone really listens and asks about my hometown, it makes me recall more details I hadn’t thought about before.''}} \textcolor{black}{These findings highlight that older drifters have different needs for the level of control exercised by the reminiscence assistant during conversations.}

\textcolor{black}{
\textbf{F2-2: Older drifters expect the assistant can generate some materials that resonate with their memories to trigger the reminiscence.} Since the lack of photos or memorabilia, many participants want the assistant to generate some materials to help them reminisce and these materials should reflect their actual experiences.} For instance, P6 was confused when an image depicting a foreign setting instead of a familiar one, and P11 stressed, \emph{``If it doesn’t understand the features of my hometown, the generated images won’t help me reminisce well.''} However, some participants agreed that materials need not be exact replicas, as long as they evoke the right emotions. As P9 said, \emph{``The details don’t need to be exact, just similar enough.''} Four participants emphasized the importance of being able to modify materials if they differed significantly from their memories, as it enhanced their engagement and deepened the reminiscence process by recalling specific details. As P9 explained, \emph{``Modifying the images helps me dig deeper into my memories, bringing up details I hadn’t thought about in years.''} Additionally, some participants, like P10, expressed a preference for fine-grained editing options to adjust specific elements within the images. \textcolor{black}{These findings highlight the need for tools that align generated content with users’ cultural and geographic contexts while offering flexible customization to enhance both emotional and cognitive aspects of reminiscence.}

\textcolor{black}{
\textbf{F2-3: Older drifters hope the assistant can help create life storybooks for sharing memories.} Four participants expressed a strong desire for the reminiscence assistant to help organize their memories into a life storybook, enabling them to share their hometown experiences with younger generations who seldom visit their hometowns.} As P1 noted, \emph{``When I get older and can no longer organize things myself, I hope the assistant can help me by putting my stories together.''} Similarly, P5 wanted this help without burdening her family. Beyond organization function, two participants hoped the tool could enrich their memories by adding details they found difficult to express clearly. As P12 explained, \emph{``I tell it in a plain language, and it can help me by adding more details, enriching and completing my memories.''} \textcolor{black}{Additionally, several participants emphasized the value of creating tangible outputs, such as photo albums with captions, to make sharing easier with grandchildren. P9 suggested, \emph{``After organizing the memories, I hope it can be printed out, made into small photo albums with text and illustrations.''} These findings emphasize the significance of tools that can support memory preservation in order to foster intergenerational sharing and cultural connection.}

\textcolor{black}{
\textbf{F2-4: Older drifters require enhanced accessibility. }}When discussing interaction preferences, most participants found voice interaction more convenient than text input, especially participants like P6, who faced challenges with hand-eye coordination and keyboard use. \textcolor{black}{As P6 explained, \emph{``Typing is not very convenient for me, so I hope to be able to use voice input.''}} P9 also preferred both voice input and output, explaining that it would help with understanding unfamiliar words. Although some participants currently have no significant difficulties with text input or reading, they anticipated voice interaction becoming more beneficial as their eyesight declined with age. P1 noted, \emph{``When I’m in my seventies or eighties, my eyesight will get worse, so voice communication will be more convenient.''} Additionally, three participants expressed a preference for a gentler, warmer chatbot tone to enhance comfort during interactions. \textcolor{black}{P1 noted, \emph{``I hope the conversation’s tone is gentler and kinder, so it feels more natural.''} These findings suggest that the reminiscence support tool is better to be characterized by good accessibility since older drifters' limited technical abilities and physical limitations.}

\subsection{Design Strategies}
\label{DS}
Through the \textcolor{black}{formative study}, we identified five key design strategies to guide the system design and implementation in Section~\ref{RemiHaven}:
\begin{itemize}
    \item \textbf{DS1:} Allow users to configure the level of control \textcolor{black}{(a guider who is familiar with their hometowns and a listener who is interested in their hometowns)} for the reminiscence assistant based on personal preferences \textcolor{black}{(F2-1)}.
    \item \textbf{DS2:} Ensure the assistant can access the user's background information to generate conversations and materials that align with their memories \textcolor{black}{(F2-2)}.
    \item \textbf{DS3:} Enable users to modify generated content, allowing adjustments to specific areas as needed \textcolor{black}{(F2-2)}.
    \item \textbf{DS4:} Enable the assistant to supplement details, organize memories, and allow the export of life storybooks into physical formats \textcolor{black}{(F2-3)}.
    \item \textbf{DS5:} Integrate voice interaction with a gentle and friendly tone to enhance accessibility \textcolor{black}{(F2-4)}.
\end{itemize}

\section{RemiHaven: \textcolor{black}{A Reminiscence Support Tool} for Older Drifters}
\label{RemiHaven}
\subsection{System Overview}
\begin{figure*}[ht]
    \centering
    \includegraphics[width=\linewidth]{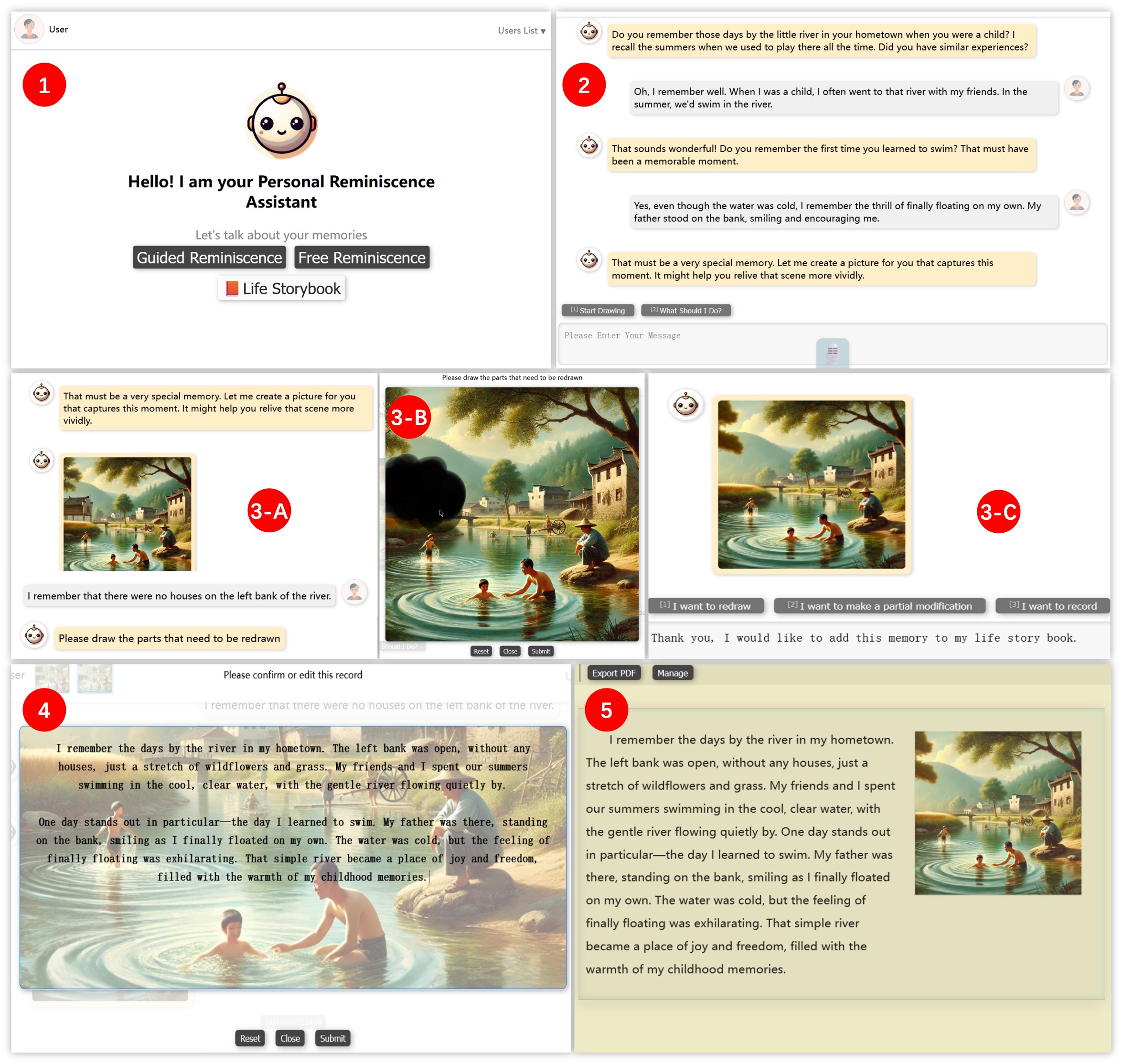}
    \caption{The RemiHaven system interfaces: (1) Users start by selecting reminiscence modes or accessing their Life Storybook. \textcolor{black}{To avoid ambiguity and make it easier for users to understand, the two modes Conversation with ``In-town'' Peers and Conversation with ``Out-of-Town'' Peers are represented in the interface as ``Guided Reminiscence'' and ``Free Reminiscence'', respectively. Before system use, we comprehensively introduced the meanings and difference of these two concepts to older drifters.} (2) The conversational interface, where the system engages users with questions to help them recall specific memories. (3-A) The initial image generated based on the user's recollections, (3-B) the image modification tool allowing users to redraw or adjust specific parts of the image, and (3-C) the finalized image ready for inclusion in the life storybook. (4) The system-generated textual description of the user's memory, which can be reviewed and edited. (5) Finally, users can export their memories as a PDF Life Storybook for saving or printing. The interface shown is translated from Chinese for clarity.}
    \Description{The RemiHaven system interface is displayed in five sections: (1) Users start by selecting reminiscence modes or accessing their Life Storybook. (2) The conversational interface, where the system engages users with questions to help them recall specific memories. (3-A) The initial image generated based on the user’s recollections, (3-B) the image modification tool allowing users to redraw or adjust specific parts of the image, and (3-C) the finalized image ready for inclusion in the Life Storybook. (4) The system-generated textual description of the user’s memory, which can be reviewed and edited. (5) Finally, users can export their memories as a PDF Life Storybook for saving or printing. The interface shown is translated from Chinese for clarity.}
    \label{fig: screenshoot}
\end{figure*}

Based on the findings and design strategies from our \textcolor{black}{formative study}, we designed and implemented a prototype system named RemiHaven, as shown in Figure~\ref{fig: screenshoot}. It is a \textcolor{black}{reminiscence support tool} powered by MLLMs to address the challenges older drifters face, particularly the difficulty in recalling meaningful memories due to a lack of tangible materials. RemiHaven guides users through reminiscence conversation, actively stimulating deeper emotional recall by generating personalized images and text that resonate with their past experiences. These multimodal outputs are ultimately compiled into a life storybook for easy saving and sharing.

\begin{figure*}[ht]
    \centering
    \includegraphics[width=\linewidth]{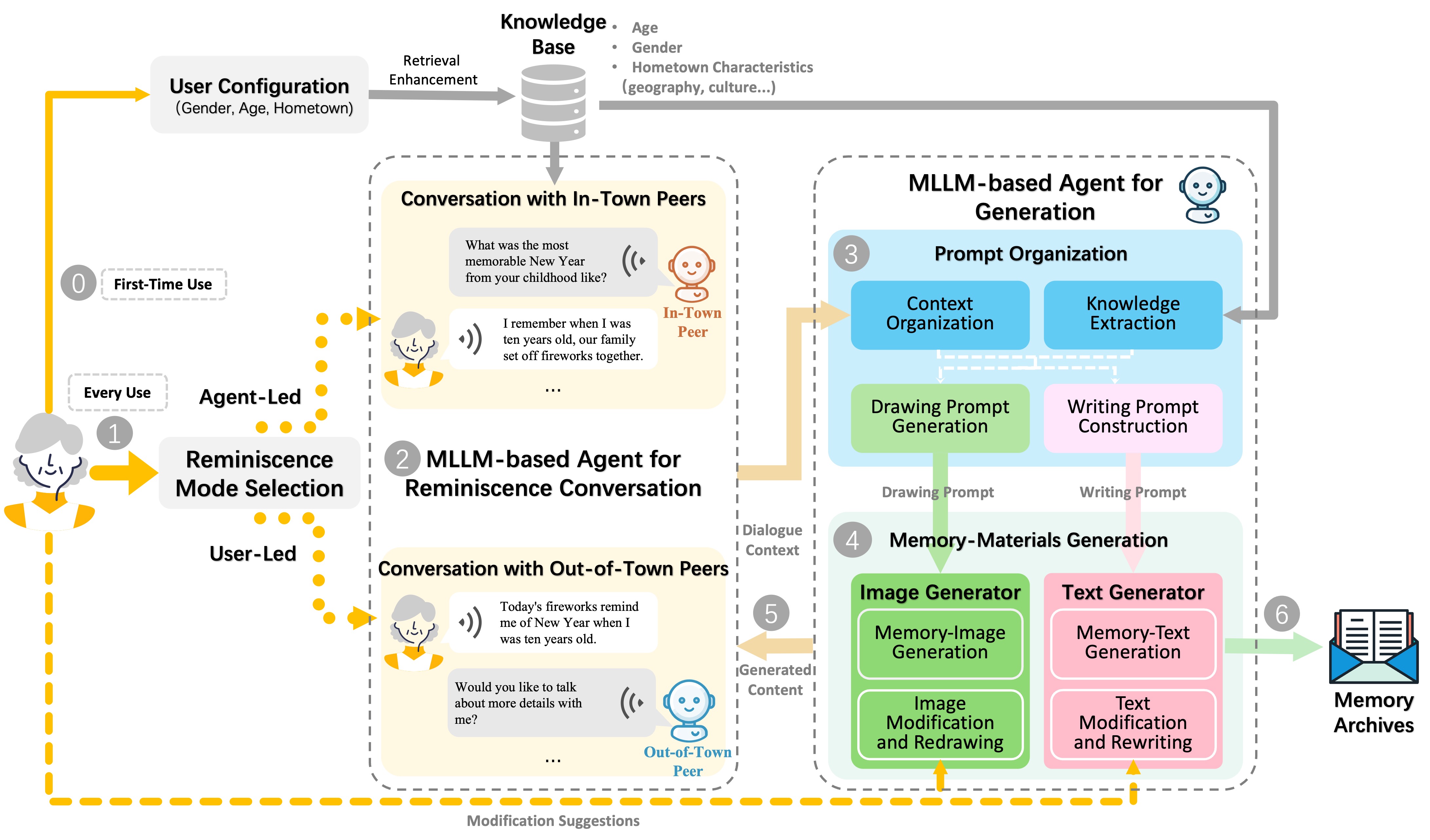}
    \caption{The architecture of RemiHaven.}
    \label{fig: architecture}
    \Description{This figure outlines the system architecture of RemiHaven, a reminiscence support tool. It begins with User Configuration, where demographic information (e.g., gender, age, hometown) is input to enhance retrieval from the Knowledge Base. Users then select either Conversation with ``In-Town'' Peers or Conversation with ``Out-of-Town'' Peers Mode. The MLLM-based Agent then organizes prompts and generates personalized Memory-Materials (images and text) in the Memory-Materials Generation phase. Users can modify the generated content, and all outputs are stored in the Memory Archives for future reference, allowing memories to be compiled into a life storybook.}
\end{figure*}

As shown in Figure~\ref{fig: architecture}, RemiHaven is composed of four main modules:
\begin{itemize}
    \item \textbf{User Configuration:} During the first use, the system requests the user’s basic information such as gender, age, and hometown (Step 0 in Figure~\ref{fig: architecture}), for generating personalized content in the latter procedures. This module supplements the user’s basic information with additional data, such as geographical knowledge and cultural background related to their hometown (Section~\ref{User Configuration}).

    \item \textbf{MLLM-based Agent for Reminiscence Conversation:} In this module, users engage in reminiscence conversations with the MLLM-powered assistant via voice or text (Step 2 in Figure~\ref{fig: architecture}). Users can choose between two modes—Conversation with ``In-Town'' Peers or Conversation with ``Out-of-Town'' Peers—depending on their preferred level of control (Step 1 in Figure~\ref{fig: architecture}). \textcolor{black}{These modes are inspired by the findings of our formative study (F2-1 descripted in Section \ref{Formative Study}) : ``In-Town'' peers simulate familiar individuals from the user’s hometown, enabling the assistant can take a more active role in guiding the conversation, while ``Out-of-Town'' peers resemble new acquaintances who show interest in the user’s hometown, letting older drifters to delve into their recollections at their own pace.} These modes offer varying levels of control to match user preference for guided or free-form interaction.

    \item \textbf{Prompt Organization:} This module organizes contextual information from the reminiscence conversations, as well as the personal background information (Step 3 in Figure~\ref{fig: architecture}) to generate prompts for creating reminiscence images and organizes the corresponding text (Section~\ref{Prompt Organization}).
    
    \item \textbf{Memory-Materials Generation:} Using prompts from the previous module, this component generates personalized images and text (Step 4 in Figure~\ref{fig: architecture}). Users can engage in deeper iterative conversations with the assistant to review, refine, and select generated content through multiple rounds of feedback (Step 5 in Figure~\ref{fig: architecture}), which is then compiled into a life storybook that can be exported for preservation and sharing (Step 6 in Figure~\ref{fig: architecture}, Section~\ref{Memory-Materials Generation}).
\end{itemize}

\subsection{User configuration}
\label{User Configuration}
Inspired by DS2 from our \textcolor{black}{formative study}, we incorporated a personalized knowledge base for each user. This allows the reminiscence assistant to understand the users' basic information during conversations, generating images and text closely aligned with their specific experiences. When a user first uses the system, RemiHaven requests their basic information (e.g., gender, age, hometown) with the user's consent, emphasizing that this information enhances the personalization of the reminiscence experience. Gender ensures that generated images and narratives reflect the user’s identity, age provides historical context by placing memories in the appropriate era, and hometown details offer cultural relevance for a more personalized experience. Our formative study highlighted that older drifters' reminiscence content is often tied to their hometowns \textcolor{black}{(F1-1)}. To enhance this, RemiHaven retrieves additional details from Baidu Encyclopedia \footnote{Baidu Encyclopedia: https://baike.baidu.com}, such as geography, cultural customs, and landmarks, integrating these into the MLLM prompts to support both the Conversation with ``In-Town'' Peers mode and the generation of reminiscence materials.

\subsection{MLLM-based Agent for Reminiscence Conversation}
\label{LLM-based Agent for Reminiscence Conversation}

\textcolor{black}{As suggested in our formative study (F1-2 depicted in Section \ref{Formative Study}), the disruption of social networks often hinders older drifters' opportunities for reminiscence conversations. To address this problem, we designed a reminiscence assistant powered by MLLMs to facilitate more engaging conversations. } Considering the varying needs for the assistant's level of control in different reminiscence scenarios (DS1), we designed two modes for reminiscence conversations: Conversation with ``In-Town'' Peers and Conversation with ``Out-of-Town'' Peers. \textcolor{black}{This design also echos the previous findings that communicating with peers can enhance older adults' emotional and cognitive proximity \cite{soon2006talking, boz2018social, underwood2010interactive}.} Before each conversation, users can select the mode that best suits their needs and preferences. To improve accessibility (DS5), we incorporated the voice interaction function, allowing older drifters to communicate with the assistant easily and naturally.

\subsubsection{Conversation with ``In-Town'' Peers}
\label{Conversation with In-town Peers}
When users lack a specific memory topic or are unsure how to start, the Conversation with ``In-Town'' Peers mode provides guidance. The assistant, acting as a peer from the user’s hometown, utilizes the user’s background information and local knowledge to evoke emotional resonance. \textcolor{black}{Based on familiar cultural and environmental elements, this mode helps users reconnect with memories.} We meticulously designed the prompts for the MLLM \textcolor{black}{(Appendix~\ref{Prompts}, Figure~\ref{fig:prompt1})}, incorporating the user's background from the knowledge base and referencing themes proven effective in reminiscence therapy, such as childhood, family, school life, festivals, and events \cite{acsiret2018effect, acsiret2018effectof} to help users recall specific memories. For example, the assistant might ask, \textit{``Do you remember the river in your hometown? I used to fish there as a child. Do you have similar memories?''} These personalized questions help users recall specific locations or events while evoking broader memories of their hometown’s landscape and childhood experiences. Throughout the conversation, the system dynamically adapts its questions based on the user’s responses and background. For instance, if a user mentions a festival, the assistant may ask about specific participants, activities, or emotions related to that event. This interactive process guides users until enough content has been gathered or they choose to stop and generate memory materials, ensuring a rich and personalized reminiscence experience. 

\subsubsection{Conversation with ``Out-of-Town'' Peers}
\label{Conversation with Out-of-town Peers}
For users with a specific memory topic or a preference for more freedom, the Conversation with ``Out-of-Town'' Peers mode is more suitable. In this mode, the assistant, portrayed as a peer from a different location but deeply interested in the user’s hometown, primarily serves as a listener. \textcolor{black}{Unlike casual conversations with neighbors or community members, this mode leverages open-ended questions to elicit personal stories and reflections that are often overlooked in typical social interactions. For instance, neighbors may lack the curiosity or shared context to sustain meaningful reminiscence, while the ``Out-of-Town'' peer expresses genuine interest in the user’s unique experiences, fostering a sense of validation and encouraging further exploration.} We designed prompts for the MLLM that emphasize openness and non-directive questioning, encouraging users to share memories at their own pace \textcolor{black}{(Appendix~\ref{Prompts}, Figure~\ref{fig:prompt2})}. For example, when a user recalls an event, the assistant might ask, \textit{``Your hometown sounds fascinating; could you tell me more about what it was like back then?''} \textcolor{black}{Such open-ended questions can help users explore memories that are often overlooked or hard to articulate, providing a supportive environment for recalling and organizing life stories. By focusing on encouraging spontaneous storytelling and avoiding disturbing, the ``Out-of-Town'' peer can provide a reminiscence-oriented interaction that goes beyond small talk.}

\subsection{Prompt Organization}
\label{Prompt Organization}
Older drifters have a strong need for an intuitive \textcolor{black}{reminiscence support tool} \textcolor{black}{(F1-3)}. To meet this, RemiHaven provides a function to generate reminiscence materials based on conversation context and the user’s background. Once the reminiscence assistant gathers sufficient memory details, it will proactively ask the user if they would like to generate these materials or the user can request the generation after multiple conversation rounds.

We use MLLMs to create high-quality prompts that ensure the generated materials accurately reflect the user’s memories \textcolor{black}{(Appendix~\ref{Prompts}, Figure~\ref{fig:prompt3} and Figure~\ref{fig:prompt4})}. Both image and text generation procedures follow a similar process by aggregating contextual information from the conversation and retrieving relevant data from the user’s knowledge base. This combines the user’s background, specific events, and details mentioned during the conversation to create comprehensive prompts. For example, if a user recalls a specific festival from their childhood, the prompt includes relevant cultural and environmental details to ensure the generated content accurately reflects the user’s memories. This multimodal approach ensures that both the visual and textual outputs resonate deeply with the user’s personal experiences.

\subsection{Memory-Materials Generation}
\label{Memory-Materials Generation}
The Memory-Materials Generation module creates both images and text, and also allows users to modify and refine the content to better align with their memories, which can trigger deeper discussions. These outputs are ultimately compiled into an exportable life storybook for preservation and sharing (DS3 and DS4).

Memory materials are generated using prompts from the previous stage, with images created through a text-to-image model and text generated by MLLMs. These are displayed in the conversation interface for review. To accommodate different user needs (DS3), the system offers editing tools. For images, users can select and adjust specific parts using a brush tool or choose to redraw the whole image. All images from a single reminiscence session are displayed at the top of the chat interface, allowing the user to select the most satisfactory one. Text can also be edited to add or refine details. This interactive process lets users customize both the visual and written components, ensuring a personal and accurate reflection of their memories. Finally, the images and text are stored in the user’s life storybook, which can be reviewed, modified, or deleted. The life storybook can also be exported as a PDF for printing and preservation.

\subsection{Implementation}
The front-end interactive web application of RemiHaven is built with HTML5 and hosted on a Python-based HTTP server. Its web-based design ensures compatibility across desktops, laptops, tablets, and smartphones, minimizing device limitations and allowing for flexible and accessible usage across multiple platforms.

For voice interaction, the system uses iFlytek’s speech recognition technology, a leading provider in China known for its high accuracy. On the voice output side, it employs iFlytek’s text-to-speech model, producing natural and friendly speech in an elderly tone, to enhance the user experience. To balance generation quality and response speed, RemiHaven uses the advanced ChatGPT-4o API for both conversation and prompt generation, along with streaming functionality for smooth interaction. For image generation, the system uses the DALL·E 3 API. However, since the editing API for DALL·E 3 is not yet available, RemiHaven relies on DALL·E 2 for editing (the performance is slightly inferior to that of DALL·E 3). \textcolor{black}{Moreover, all prompts used in the RemiHaven are crafted based on OpenAI’s official guidelines \cite{openai_prompt_engineering} and iteratively refined by the first author.}

All functions in the system are designed with a modular approach, ensuring each module operates independently with clearly defined interfaces. This simplifies future upgrades and feature expansions, allowing quick adaptation to new technological developments and enhancing both maintainability and scalability.

\subsection{Characteristic Summarization}
In summary, guided by its design strategies, RemiHaven possesses the following notable features:
\begin{itemize}
    \item \textbf{Personalized Reminiscence Experience (DS1, DS2):} RemiHaven generates reminiscence materials matched to the user’s personalized needs and offers multiple reminiscence modes. The User Configuration module collects and enhances the user’s background information for personalization. Users can choose between Conversations with ``In-Town'' Peers and ``Out-of-Town'' Peers modes, offering flexibility based on their preferences for a personalized experience.
    
    \item \textbf{Multimodal Reminiscence Material Generation (DS4):} The system generates multimodal reminiscence materials, including images and text, through the Prompt Organization and Memory-Materials Generation modules. Users can modify and select these materials to compile a life storybook, which can be exported as a PDF for printing and preservation. 
    
    \item \textbf{Enhanced Interactivity (DS3, DS5):} To address the challenges older drifters may face with technology, RemiHaven emphasizes user-friendly interactivity tailored to elderly users. It supports intuitive voice input and uses a text-to-speech model for natural and friendly speech. Users can also modify or redraw specific image parts with a brush tool, enhancing ease of use and improving the overall experience.
    
\end{itemize}

\section{EVALUATION}
\label{Evaluation}

\textcolor{black}{\subsection{Method}}
\subsubsection{Participants}
We recruited 10 older drifters (7 women and 3 men), aged between 62 and 78, for our evaluation. Eight participants (P2, P5, P6, P8, P9, P11, P12, and P13) were from our formative study, who expressed interest in participating our further investigation. To enhance the evaluation, we recruited two additional participants through community outreach efforts: P14, a 66–70-year-old woman from Henan, who relocated 8 years ago, and P15, a 61–65-year-old woman from Anhui, who has lived away from her hometown for 10 years. This supplementing enhanced the diversity of participants in the evaluation.

\subsubsection{\textcolor{black}{Procedure}}
\textcolor{black}{To comprehensively evaluate RemiHaven, we employed a mixed-method approach that combined quantitative and qualitative methods and adopted two sets of metrics: the Technology Acceptance Model (TAM) \cite{marangunic2015technology} and the system’s alignment with design strategies. }
The TAM framework examines users’ perceived usefulness (PU), perceived ease of use (PEOU), attitude toward using (ATU), and behavioral intention to use (BIU), \textcolor{black}{which can offer a set of structured measures to understand users' overall technology acceptance of RemiHaven \cite{davis1989perceived, nov2008users}}. Drawing upon established studies of TAM \cite{davis1989perceived}, we designed a questionnaire using a 5-point Likert scale, and the specific indicators and corresponding items are detailed in Figure~\ref{fig: result2}. \textcolor{black}{Additionally, we also assessed the system’s alignment with our design strategies, i.e., whether the system met the design expectations of older drifters in practical use.} The corresponding metrics are shown in Table~\ref{tab: metrics} and we also collected participants' feedback through questionnaires (detailed in Figure~\ref{fig: result1}).

\begin{table*}[ht] \small
\centering
\caption{Design strategies and corresponding evaluation metrics.}
\label{tab: metrics}
\begin{tabular}{>{\centering}m{2.5cm} m{4.5cm} m{6.5cm}}
\hline
\textbf{Design Strategies} & \textbf{Description} & \textbf{Metric} \\ \hline
DS1 & Allowing users to configure the level of control for the reminiscence assistant & 1a. Configuration Flexibility \newline 1b. Satisfaction with Conversation with ``In-Town'' Peers \newline 1c. Satisfaction with Conversation with ``Out-of-Town'' Peers  \\ \hline
DS2 & Enhancing the system’s knowledge about users & 2a. Background Knowledge Accuracy \newline 2b. Generated Content Relevance \\ \hline
DS3 & Offering fine-grained modification features for generated content & 3a. Modification Function Ease of Use \newline 3b. Modification Function Effectiveness \\ \hline
DS4 & Assisting users in supplementing memory details and generating a life storybook & 4a. Detail Supplementation Satisfaction \newline 4b. Memory Archive Generation Quality \newline 4c. Memory Archive Export Satisfaction \\ \hline
DS5 & Optimizing the experience through voice interaction & 5a. Voice Input Ease of Use \newline 5b. Voice Output Naturalness \newline 5c. Reminiscence Assistant Tone Satisfaction \\ \hline
\end{tabular}
\end{table*}

\textcolor{black}{To complement these quantitative measures, we incorporated qualitative analysis to deeply explore user experiences, e.g., RemiHaven's overall strengths and weaknesses to be improved.} During the system use, participants employed the think-aloud protocol \cite{charters2003use}, articulating their thoughts and reactions in real-time while engaging with the system. Then participants discussed both the system’s strengths and areas for improvement in the subsequent semi-structured interviews. 
\textcolor{black}{Furthermore, participants' interaction data was recorded with their consent to enhance our findings, including their screen activities and audio inputs.}

\begin{figure}[ht]
    \centering
    \includegraphics[width=\linewidth]{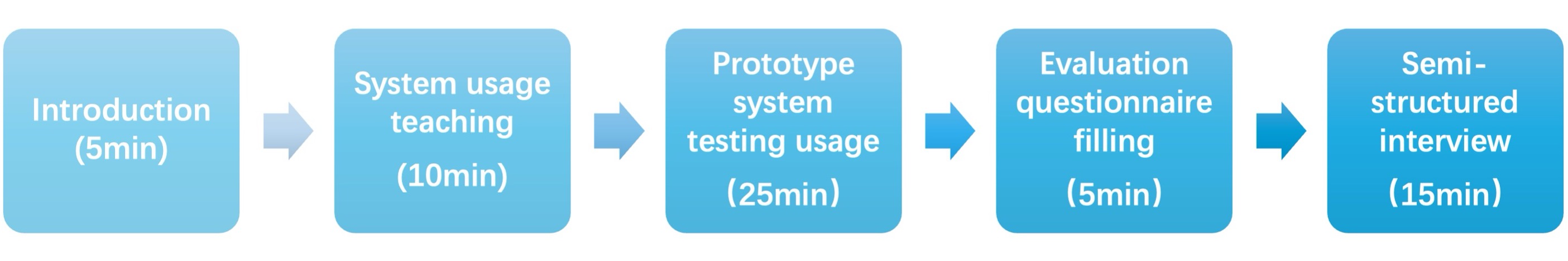}
    \caption{The evaluation procedure.}
    \Description{The diagram illustrates the step-by-step process for evaluating the system, starting with a 5-minute introduction, followed by a 10-minute teaching session on system usage. Participants then spend 25 minutes testing the prototype system, after which they complete a 5-minute evaluation questionnaire. The process concludes with a 15-minute semi-structured interview to gather in-depth feedback.}
    \label{fig: evaluation}
    \end{figure}

Evaluations were conducted in community centers, following the steps in Figure~\ref{fig: evaluation}. Participants first received detailed explanations of the evaluation process, necessary precautions, and an introduction to the prototype's features and operation. They then selected their preferred reminiscence mode and engaged in 1-2 rounds of reminiscence using RemiHaven on a PC, with researcher's assistance. During the trial, we closely monitored interactions and requested participants to use the ``think aloud'' method \cite{charters2003use} to provide real-time feedback. Afterward, they completed a questionnaire that corresponded to the two parts of the evaluation metrics using a 5-point Likert scale, followed by semi-structured interviews to gain deeper insights into their overall impressions of the system and expectations for further development. Each participant received a compensation of 50 yuan after the interview, and all procedures followed ethical guidelines outlined in Section~\ref{ethical}.

All Recordings were transcribed using automated tools and reviewed by the first author. Quantitative data from the questionnaires were analyzed using statistical methods. We calculated the mean scores for each question to evaluate system features and user experiences, with reliability confirmed through Cronbach’s Alpha. All statistical analyses were performed using SPSS. For qualitative data, thematic analysis was applied to the interview transcripts, following a structured coding process similar to our formative study. \textcolor{black}{We also analyzed user behaviors from the interaction data, such as the frequency, duration, and operational patterns of using each feature, which served as supplementary evidence for our quantitative and qualitative findings.}

\subsection{Result}
\begin{figure*}[ht]
\centering
\includegraphics[width=\linewidth]{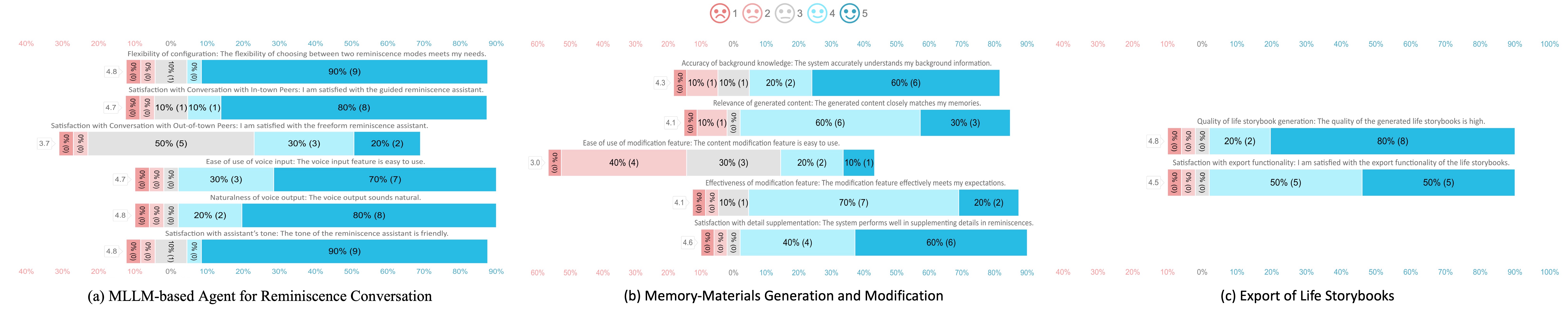}
\caption{Quantitative evaluation results in terms of functionality. Blue represents a score of 4 (satisfaction) or 5 (strong satisfaction); red represents a score of 1 (strong dissatisfaction) or 2 (dissatisfaction); and gray represents a score of 3 (neutrality).}
\Description{The figure presents the evaluation results across three areas: (a) MLLM-based Agent for Reminiscence Conversation, (b) Memory-Materials Generation and Modification, and (c) Export of Life Storybooks. Each section shows the satisfaction levels of participants using a 5-point Likert scale, with most participants expressing high satisfaction (scores of 4 and 5) across categories like conversation flexibility, accuracy of generated materials, ease of modification, and the quality of life storybooks. Notably, the flexibility of reminiscence modes and the export functionality received the highest scores, reflecting strong user approval.}
\label{fig: result1}
\end{figure*}

\begin{figure*}[ht]
\centering
\includegraphics[width=0.8\linewidth]{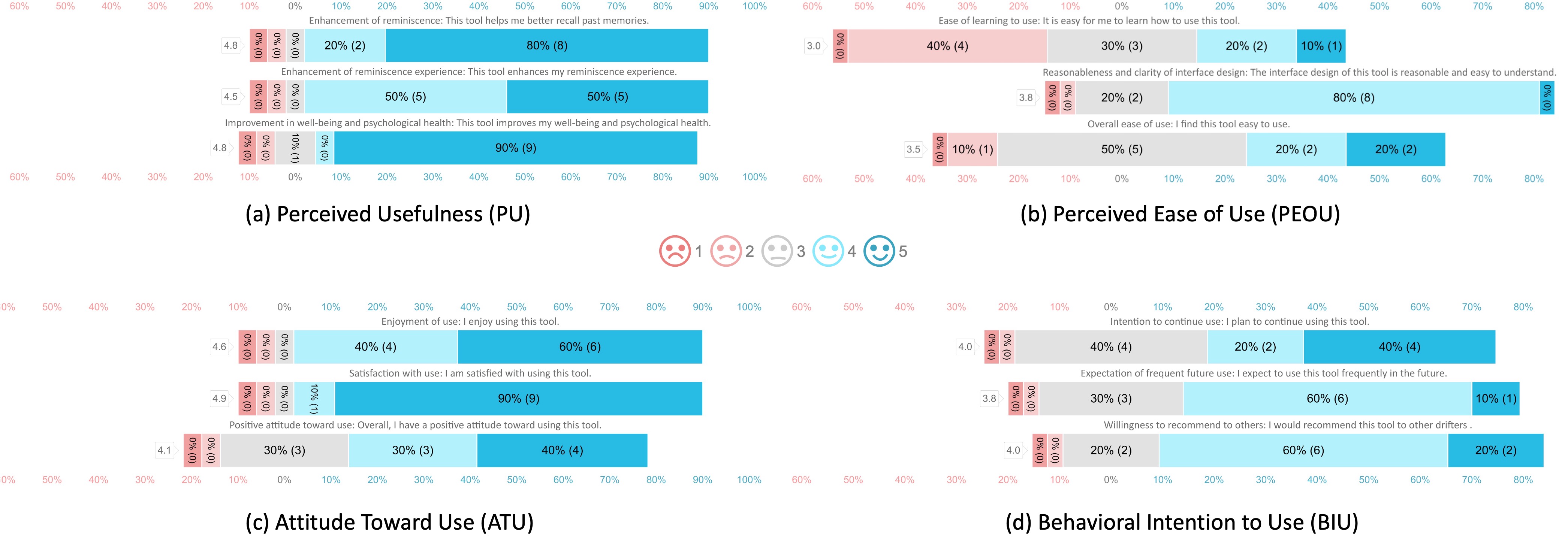}
\caption{Quantitative evaluation results from the overall perspective (using the same color representation as Figure~\ref{fig: result1}). }
\Description{The figure illustrates the evaluation of the system based on four key dimensions: (a) Perceived Usefulness (PU), (b) Perceived Ease of Use (PEOU), (c) Attitude Toward Use (ATU), and (d) Behavioral Intention to Use (BIU). Each dimension is rated on a 5-point Likert scale, with high scores (4 and 5) dominating across all categories. Notably, the system scored highly in perceived usefulness. Similarly, most participants found the system easy to use (PEOU) and expressed a positive attitude towards its continued use (ATU), indicating strong user acceptance and intention to use the tool in the future (BIU).}
\label{fig: result2}
\end{figure*}

During the evaluation, each participant completed a 20-30 minutes session using the system, which covered all core functionalities of our prototype. We assessed the reliability of the questionnaire data with Cronbach’s coefficient alpha \cite{cho2015cronbach} and achieved a score of \(\alpha = 0.925\), indicating strong reliability. Evaluation results were organized by RemiHaven's main functionalities (Figures~\ref{fig: result1} and ~\ref{fig: result2}). Combining quantitative data with qualitative insights from the semi-structured interviews \textcolor{black}{and interaction data analysis}, we identified the following key findings.

\subsubsection{MLLM-based Agent for Reminiscence Conversation}
\label{eva-1}
Firstly, we evaluated the performance of the MLLM-based reminiscence assistant during reminiscence conversations. Analysis of the questionnaire data (as shown in Figure~\ref{fig: result1}-a), interview transcripts, and interaction records showed overall positive feedback from participants across several aspects, while certain areas still require improvement.

\textbf{The flexible reminiscence options effectively met participants' personalized needs.} Participants generally felt that both modes could meet their reminiscence needs. Data shows that 90.0\% of participants rated the flexibility of the reminiscence modes at 4 or higher, with an average score of 4.8, indicating RemiHaven's adaptability to different scenarios and personalized needs. For instance, P13 noted, \emph{``Sometimes I like to talk freely, and sometimes I feel that being guided makes it easier to reminisce, so both modes work well for me.''} This feedback further confirms the system's flexibility in accommodating various reminiscence needs and scenarios.

\textbf{Participants' feedback varies across the two agent roles.} When comparing the Conversation with ``In-Town'' Peers and Conversation with ``Out-of-Town'' Peers modes, participants generally preferred the former, with 90.0\% rating their satisfaction with it as 4 or higher, and the average score is 4.3. In contrast, the Conversation with ``Out-of-Town'' Peers mode received a slightly lower average score of 3.7. \textcolor{black}{Interaction record also echos this finding, i.e., participants spent an average of 16 minutes in the Conversation with ``In-Town'' Peers mode but 12 minutes in the Conversation with ``Out-of-Town'' Peers mode. It is likely that guided conversations of ``In-Town'' peers can result in higher engagement compared with ``Out-of-Town'' peers.} 

This difference may stem from participants' stronger identification with familiar local backgrounds and shared experiences. Conversations with peers from the same hometown often involve familiar settings, making it easier for participants to resonate with the content and feel that the dialogue is more intimate and natural. However, in the Conversation with ``Out-of-Town'' Peers mode, participants' thoughts tended to be more dynamic during reminiscence, posing greater challenges for the system's understanding and response. Some participants reported that the system struggled to capture and respond to their shifting thoughts, affecting the conversation's coherence. For instance, P13 mentioned that when she tried to interrupt the system to switch to a new topic, the system lagged behind, negatively impacting her experience: \emph{``When I wanted to start recalling something else, it was still talking about the previous topic. By the time it did, I forgot what I was going to say.''} Additionally, we also observed \textcolor{black}{from screen records} that the assistant sometimes struggled to balance listening and questioning, occasionally interrupting users’ reminiscence flow. This issue may be related to the current dialogue structure, where MLLMs are inherently more inclined towards expression rather than listening, and users can only input again after the model outputs a complete response. Although this design helps maintain the coherence of the dialogue, it can be less flexible in situations where users seek more free-form interaction. Future improvements should focus on better aligning the system with user needs across different modes to create more natural interactions.

\textbf{The voice interaction mode promotes system usability, while some improvements are needed.} The voice interaction function received overall positive feedback, with 100\% of participants rating the voice input feature 4 or higher. This feature was particularly beneficial for users with limited literacy, such as P14, who noted, \emph{``Even though I don't recognize characters, I can still chat with it, just like talking face-to-face with a real person.''} Additionally, voice interaction reduced older drifters' potential interaction barriers resulted by age-related vision issues \cite{zeng2016vision} and decline in text processing ability \cite{nicolau2012elderly}. \textcolor{black}{Interaction data further highlighted the effectiveness of voice input, i.e., users tended to rely on voice interaction for longer expressions. However, for participants with less standard Mandarin pronunciation, the system sometimes misinterpreted commands, leading to repeated attempts and minor frustration.} 

While the majority of participants were satisfied with the voice output (90.0\% giving high scores for the tone and the elderly female voice that we set), a few participants, like P15, felt that the voice sounded ``too old'' and suggested offering younger-sounding options. P15 remarked, \emph{``A younger voice would make me feel younger as well.''} This indicates opportunities for improvement by introducing more diverse voice selections to better align with individual preferences. Some participants also suggested incorporating dialect support to enhance familiarity, making the system more aligned with users' everyday language habits. For example, P9 remarked, \emph{``If it spoke in our hometown dialect, it would feel much more familiar.''}

\textbf{Users expressed a desire for more input modalities to improve their interaction experience.} During the evaluation, three participants expressed a desire for multimodal inputs, such as images and videos, to assist with reminiscence, believing it could help the assistant better understand the conversation content and stimulate deeper memories. \textcolor{black}{The corresponding screen records also indicate they spent more time verbally describing visual elements they wished to share.} For example, P8, an engineer, wanted to use hand-drawn sketches to illustrate his work environment but was disappointed that the system did not support image input: \emph{``When I chat with others, I like to draw things out, but since it can't see, I find it hard to explain clearly.''} Additionally, P12, while using the Conversation with ``In-Town'' Peers mode, noted that the assistant showed interest in her hometown. She expressed a strong desire to share photos with the assistant. Incorporating such multimodal inputs could not only enrich the ways users express themselves but also help the system more accurately understand their intentions, thereby enhancing the overall reminiscence experience. 

\subsubsection{Memory-Materials Generation and Modification}
\label{eva-2}
Analysis of the questionnaire data (as shown in Figure~\ref{fig: result1}-b), interview transcripts, and interaction records revealed that this module was overall appreciated by participants. However, there remains opportunities for improvement in certain aspects.

\textbf{The image generation feature effectively facilitated reminiscence, while the quality of generated images should be considered from multiple dimensions.} Participants generally felt that the generated images effectively triggered their memories, with some even stating that seeing these images made them feel as if they were reliving in the past: \emph{``It felt like everything happened just yesterday.''} 

When we asked participants about their satisfaction with the generated images, they typically evaluated the quality based on two factors: aesthetic effect and the similarity between the depicted scenes and their memories. The data shows that 90.0\% of participants rated the system's performance in image generation as 4 or higher, with an average score of 4.1. This indicates users' overall satisfaction with the generated images. Some participants noted that the images did not always accurately match their recollections, reflecting a potential gap in the system's grasp of background knowledge and biases in the image generation model. However, for most users, high similarity was not always required. As P12 mentioned: \emph{``I just need an image that reminds me of the memory; it’s unrealistic to expect it to look exactly like what’s in my mind after all these years.''} This feedback suggests that participants have varying expectations regarding the similarity between generated images and their memories. Future system designs may need to strike a balance between the artistic and realistic aspects of image generation to better meet the diverse needs of different users.

\textbf{The text generation performed well in supplementing memory details.} Participants generally reported that the generated text accurately captured their memories and effectively supplemented the details. Data shows that 80\% of participants accepted the initial output without adjustments, and 100\% of participants rated their satisfaction with the generated text at 4 or higher, with an average score of 4.6. This efficiency reflects the system’s strong alignment with user expectations. This indicates the system's text generation not only aligns well with users' memories but also provides meaningful detail enhancement, further enriching the reminiscence experience. As P9 mentioned: \emph{``The generated text helped me recall more details like it was organizing my thoughts.''} This seems that the current capabilities of MLLMs in generating text remain stronger than those in image generation.

\textbf{To better generate memory-related images and text, the MLLMs need to improve their understanding of participants' hometown backgrounds and cultural details.} Although the functions of generating images and text performed well overall, the MLLMs showed certain limitations in understanding the cultural backgrounds and folk details specific to participants' hometowns. \textcolor{black}{We observed that the system occasionally produced content that deviated from participants' memory details and users often tried to revise the inappropriate culture-related descriptions.} For instance, P2 recalled a traditional game from her hometown, but the system generated an image that significantly differed from her memory, reflecting the system's insufficient grasp of certain regional cultures and folklore. P2 remarked: \emph{``The picture is nothing like the game we used to play as kids; it feels like it was just randomly drawn.''} This suggests a need for MLLMs to further enhance their understanding and integration of background knowledge when handling specific cultural contexts and folk details. Future improvements could consider incorporating more extensive cultural knowledge databases.

\textbf{The content modification function enhanced users' sense of control, while its usability still needs improvement.} The image modification feature was overall appreciated, with 90.0\% of participants rating it 4 or higher and an average score of 4.1. \textcolor{black}{We found that participants typically required an average of 1.8 modification attempts to achieve a satisfactory result.} Participants mentioned that the modification feature allowed them to further adjust the generated content, making it more aligned with their memories and enhancing their sense of control over the reminiscence process. The process of modification, in conjunction with the conversation, further deepened the reminiscence. For example, P14 noted: \emph{``At first, it didn’t quite match, but after a few modifications, it got closer and even made me notice details I hadn’t thought of before.''}

However, some participants encountered difficulties when using the modification function, particularly those who were less familiar with the technology. P12 mentioned: \emph{``Modifying it was a bit complicated; sometimes I wasn’t sure how to make the changes.''} This indicates that while the modification feature provides significant flexibility, its usability still needs improvement.

\subsubsection{Export of Life Storybooks}
Through the analysis of questionnaire data (as shown in Figure~\ref{fig: result1}-c), interview transcripts, and interaction records, we found that the generation and export functions of the life storybooks were overall appreciated by participants, particularly for preserving and sharing personal memories.

\textbf{The generated life storybooks received positive feedback from participants and held significant emotional value.} In the interviews, several participants mentioned that the life storybooks were not only a record of personal memories but also served as a means of self-continuity (i.e., a sense of connection between one’s past and one’s present). All participants rated the quality of the life storybooks at 4 or higher, with an average score of 4.8, indicating they were highly recognized for their visualization and content presentation. For example, P6 commented: \emph{``I really like the combination of a picture with a piece of text; it's a great way to record my memories.''} P9 added: \emph{``I could recall the past without this system, but with the life storybook, these memories can be preserved longer and shared with my children.''} This feedback highlights the life storybooks' value for both older drifters themselves and memory sharing across generations.

\textbf{The export function increased the practicality of life storybooks, making them easier to preserve and share.} Participants generally felt that the export feature significantly improved their ability to preserve and share these precious memories. All participants rated the quality of life storybooks at 4 or higher, with an average score of 4.5. P9 mentioned: \emph{``Printing it out makes it much easier to revisit when I want to reminisce.''} This feature not only increased the practical value of the life storybooks but also made it easier for participants to manage and share their memories. P15 stated: \emph{``I like to save the ones I love; I enjoy collecting them.''} This feedback indicates that the export function turns the life storybooks from mere memory records into ``emotional assets'' that can be passed down and treasured. Future system designs could further optimize the export options to meet users' needs in different contexts.

\textbf{The structure and coherence of the life storybooks could be further optimized.} Although most participants were satisfied with the generated life storybooks, some desired more structured content, beyond just a simple combination of images and text. For example, P9 mentioned: \emph{``I want to record my entire experience in a sequence, like a complete story, rather than just these scattered images and text.''} This feedback reflects participants' desire for the system to provide more logically and narratively organized content to better represent their life journeys. By systematically organizing reminiscence content by time or theme, future system design could make a storybook not just collection of memories but realistic reflection of life stories.

\subsubsection{Overall}
Through the analysis of questionnaire data, interview transcripts, and interaction records, we found that users generally held a positive attitude toward the system, acknowledging its significant impact on supporting reminiscence and \textcolor{black}{improving mood}. However, there are still some areas for future improvement.

\textbf{The system excelled in supporting reminiscence and improving mood.} In terms of PU, participants generally agreed that RemiHaven significantly aided their reminiscence process. As shown in Figure~\ref{fig: result2}-a, 100\% of participants felt the system improved their reminiscence, with average scores of 4.8 for aiding memory recall and 4.5 for overall reminiscence experience. Additionally, 90.0\% of participants believed that using RemiHaven increased their perceived happiness and \textcolor{black}{fostered positive emotional responses}, with an average score of 4.8. In the interviews, several participants mentioned that the system not only helped them recall past events more clearly but also brought about positive emotional effects. For example, P2 stated: \emph{``When I can't go back to my hometown, using this helps me alleviate some of my homesickness.''} \textcolor{black}{Through observing users’ emotional changes from interaction records, we noticed a positive shift from initial unfamiliarity with the system to actively sharing memories, particularly when the system accurately responded to their memories.}

Furthermore, some participants expressed that although they were eager to share their memories with others, they were concerned about taking up too much of the other person's time. However, when sharing memories with such an assistant, they had no such worries. As P12 mentioned: \emph{``If I can use this anytime without worrying about wasting someone else's time, I would reminisce about my life's journey. These experiences are like movies playing in my mind.''} This feedback aligns with the perceived usefulness dimension of the TAM model.

\textbf{The interface of the system was well accepted, but the usability needs to be improved.} In terms of PEOU, as shown in Figure~\ref{fig: result2}-b, 80.0\% of participants felt that the system's interface design was reasonable and easy to understand, with an average score of 3.8, indicating that the system's intuitive design and usability were recognized by users. However, only 40.0\% of participants rated the overall usability at 4 or higher, with an average score of 3.6, suggesting that there is still room for improvement. Some participants encountered difficulties during the trial, which may be related to the PC we used. The unfamiliarity with computers among most older drifters led to operational challenges. For example, P12 mentioned: \emph{``We don't really use computers; those are for the younger ones.''} Future development should consider deploying the system on devices more familiar to older users, such as smartphones or televisions, to better align with their usage habits. 

Additionally, 40.0\% of participants found it somewhat difficult to learn how to use the tool, with an average score of 3.0. P12 further pointed out: \emph{``For someone like me with a lower education level, learning how to use it is a bit challenging.''} While the design is generally user-friendly, optimizing user guidance, simplifying the process, and reducing the learning burden are important for older users with lower ICT literacy.

\textbf{Users' attitudes towards the system were generally positive, with strong intentions for future use.} In terms of ATU and BIOU, participants generally held a positive attitude towards RemiHaven. As shown in Figure~\ref{fig: result2}-c, 100\% of participants reported that they liked using the system and were satisfied with their overall experience, with average scores of 4.6 and 4.9, respectively. This indicates a high level of user approval during actual use. Additionally, 70.0\% of participants anticipated further system improvements, suggesting that users have high expectations for the system's future potential. Participants widely believed that RemiHaven has significant potential for supporting reminiscence among older drifters. Most participants indicated that they would continue to use RemiHaven in the future and would recommend it to others. As shown in Figure~\ref{fig: result2}-d, the average scores for three related metrics were 4.0, 3.8, and 4.0, respectively. P5 mentioned: \emph{``Once you further refine the system, I would love to use it to complete my life storybook and share it with every member of my family.''} This feedback indicates that RemiHaven not only received positive evaluations at its current stage but also has the potential for broader application and promotion in the future.

\textbf{Summary.} In conclusion, the system demonstrated significant advantages in enhancing the reminiscence experience and \textcolor{black}{contributing to emotional benefits.} Moreover, the majority of users expressed a positive attitude towards the system and its future potential. However, the evaluation also highlights areas for future improvement, including the need to support more diverse input and output modalities, optimize the structured generation of life storybooks, enhance system usability, and reduce the learning burden.

\section{DISCUSSION}
\label{Discussion}
Our evaluation results indicate RemiHaven's potential in supporting reminiscence for older drifters. The system not only effectively stimulates users' memories but also plays a positive role in enhancing the reminiscence experience and \textcolor{black}{contributing to emotional benefits.} With its different conversation modes, RemiHaven shows flexibility in addressing personalized needs and adapting to diverse usage scenarios. However, several critical issues require further exploration, such as striking a balance between user control and LLM control and alleviating the bias of generation. The following discusses these findings and problems in detail.

\subsection{Cons and Pros of ``In-Town'' and ``Out-of-Town'' Peers in Supporting Reminiscence Conversations}

Our evaluation indicates that the introduction of ``In-Town'' peers and ``Out-of-Town'' peers enhances RemiHaven's capacity to meet older drifters' personalized needs. Inspired by the previous findings that LLM-powered Agents can significantly impact user interactions by adjusting background, personality, and interaction styles \cite{liu2024peergpt, lu2023miracle}, we integrated the two characters into RemiHaven to improve the efficiency and user experience of reminiscence for older drifters. According to our evaluation, ``In-Town'' peers effectively help users recall familiar scenes and experiences from their hometowns, as older adults prefer to communicate with partners who share similar cultural backgrounds. This design can not only alleviate communication barriers \cite{soon2006talking, boz2018social} but also foster deeper interaction through reminiscing about past events \cite{underwood2010interactive}. Older drifters in our interviews noted that speaking with a ``peer'' from their hometown felt like chatting with an old friend, making it easier to recall detailed memories. In contrast, ``Out-of-Town'' peers offer a more open mode of reminiscence, encouraging older drifters to share and reflect on their past from different perspectives. Our evaluation showed the introduction of these two characters offers more tailored reminiscence experiences. 

However, as noted in Section~\ref{eva-1}, current LLMs face limitations in supporting these characters. The most prominent issue is that LLMs struggle with adapting to more open-ended reminiscence processes, especially when older drifters rapidly change thoughts and topics during Conversations with ``Out-of-Town'' Peers. This challenge may arise from existing LLMs' tendency to maintain conversational continuity \cite{liang2024c5}, which can diminish flexibility in free reminiscence. Additionally, the target of most current AI research and development is to train LLMs or other AI models as a powerful and knowledgeable agent, making LLMs tend to dominate a conversation in practice, potentially hindering the user's initiative in the conversation. Therefore, achieving a balance between user control and LLM control is crucial in reminiscence conversations. In reminiscence scenarios, users typically prefer to have more control, enabling them to freely switch topics and guide the conversation, while LLMs should primarily act as a listener and supporter at this time. In related fields, many recent studies debate the boundaries of AI control versus human autonomy \cite{sankaran2021exploring, calvo2020supporting}. For example, in emotional support scenarios, AI's excessive intervention in the user's thinking process, rather than listening, can lead to discomfort \cite{laitinen2021ai}. Similarly, in creative domains, AI's over-dominance may stifle creativity and limit users' freedom of expression \cite{larsson2022towards}. Thus, a promising further research topic is to explore how to incorporate more flexible control mechanisms to allow the powerful AI to better adapt to the user's cognitive flow while maintaining conversation coherence and interactivity in special scenarios like reminiscence and emotional support.

\subsection{The Trade-off between Reminiscence Process and Reminiscence Archive}
In RemiHaven's reminiscence support, the generative capabilities of MLLMs are essential. The images and text generated by MLLMs can not only fill memory gaps but also promote emotional responses and the overall efficacy of reminiscence. However, during the evaluation, we found there is a trade-off between older drifters' requirement for image fidelity during reminiscence and that of the final product, giving rise to a potentially important research topic of how to balance the content quality of the reminiscence process and the final archive.

First, we found that the generated images during the reminiscence process play a crucial role in promoting users' memory recall (Section~\ref{eva-2}). The reminiscence process holds significant value, as it is dynamic and gradual. With RemiHaven, older drifters can progressively recall more details by adjusting and modifying MLLM-generated images that may not resemble their memories. This ``from dissimilar to similar'' process is a key mechanism for older drifters to structure their memories and expand the scope of reminiscences. On one hand, if the generated content misaligns with users' memories, particularly when cultural or personal elements are misrepresented or overlooked, it may lead to users' confusion and hinder effective reminiscence triggers \cite{struppek2022biased}. On the other hand, we found high-fidelity content may provoke negative emotional reactions, such as the ``uncanny valley'' effect, indicating images that are highly similar to reality but unnatural may cause discomfort or fear \cite{macdorman2009too}. Additionally, when the generated images appear too realistic, older drifters might feel like their inner world is being ``invaded'', leading to privacy concerns and potential resistance to the system \cite{chen2023challenges, amer2023ai}. Thus, we thought in the Generative AI-supported reminiscence process of older drifters, the most important is their perception rather than the generated content's high intrinsic quality. Such system design should pay more attention to how to generate content properly and gradually to ensure it can appropriately invoke older drifters' memories step-by-step without fear and worry.

Second, we found when shifting from the reminiscence process to creating a final product, older drifters' expectations for image quality change a lot. At this stage, older drifters tend to obtain high-fidelity images to facilitate sharing with family and friends. High-fidelity images can enhance the vividness and preservation of memories, playing a key role in storytelling \cite{agroudy2016impact}. Considering the conflict of older drifters' expectations for image quality between the reminiscence process and product creation, there is a need to focus on exploring coordination strategies in future research. On the one hand, the generated content should overall align with users' memories, paying more attention to cultural context and key figures to avoid confusion. On the other hand, flexibility in artistic presentation and interpretation, such as abstract representations or symbolic elements, can be reserved, which can prevent the ``uncanny valley'' effect and encourage users' deeper reflection. Furthermore, personalization might also be a promising solution. By allowing older drifters to choose the style of image generation (realistic style, illustration style, etc.), the generated images can avoid ``over-fidelity'' and are more helpful in addressing their emotional and cognitive needs. With these strategies above, MLLMs are expected to play a more helpful role in reminiscence support in the future.

\subsection{Implications for Generative AI Enhanced Reminiscence-Based Interventions}

Our study shows that generative AI offers new possibilities for reminiscence-based interventions for older drifters. These techniques can not only promote the reminiscence experience and \textcolor{black}{contribute to emotional benefits} but also simplify intervention implementation, reducing the workload of social workers. The following summarizes the strengths of generative AI in supporting reminiscence-based interventions.

\begin{itemize}
    \item \textbf{Enhancing Reminiscence Experience and Contributing to Emotional Benefits for Older Drifters: } According to our evaluation (Section~\ref{eva-1}), the introduction of generative AI significantly enhances older drifters' engagement in reminiscence, making it easier to record and share memories. AI-generated conversations reduce feelings of loneliness by providing emotional support and connecting older drifters with their past. Additionally, the generation of memory-related content allows older drifters to vividly reflect important life moments, \textcolor{black}{contributing to positive emotional experiences. By fostering meaningful reflections and connections, the system aligns with the Flourishing Through Leisure model \cite{anderson2012flourishing}, which emphasizes recreation as the purposeful facilitation of quality leisure experiences to enhance well-being. It provides a methodology to supply individualized support for people facing life transitions, helping them achieve their goals and promote spiritual strengths.} The generated life storybooks also preserve precious memories and serve as a medium for reflection and sharing with family and friends, helping strengthen social bonds \cite{liaqat2022hint}. Moreover, RemiHaven's natural language interactions and image generation based on natural conversations reduce older drifters' barriers to adoption. This design enables users to enjoy the benefits of the system without learning specialized technical skills, reducing potential stress and anxiety associated with new technology use \cite{yusif2016older}.
    
    \item \textbf{Reducing the Workload of Human Supporters:} Generative AI in reminiscence-based interventions shows promising potential for social impact by reducing the time and effort required for social workers \cite{tam2021effectiveness}. Automating content creation and conversation facilitation increases intervention efficiency, allowing social workers to focus on more personalized support. Their collaboration can help expand reminiscence-based intervention services to a larger population beyond older drifters. \textcolor{black}{Generative AI can assist older adults} in organizing and sharing personal experiences more efficiently, helping alleviate common issues such as loneliness and cognitive decline \cite{tam2021effectiveness}. For other migrant groups, particularly children and adolescents, generative AI techniques have been utilized for recording migration experiences and family histories, fostering their understanding of own cultural background and belonging in new environments\cite{liaqat2022hint, bala2024stories, hedditch2023design, gao2022taking}. Therefore, we believe Generative AI tools hold broad potential in supporting social work for diverse communities by complementing human efforts. Future research can further optimize AI’s role across different populations to better serve societal needs.
\end{itemize}

However, despite the advantages that generative AI demonstrates in reminiscence-based interventions, its potential risks and challenges cannot be overlooked.

\begin{itemize}
   \item \textcolor{black}{\textbf{Integrating AI into Older Drifters’ Reminiscence Practices and Social Dynamics:} The integration of AI into older drifters’ reminiscence practices may affect their real-world social relationship by shifting their reliance for memory sharing and emotional support from humans to AI. In their hometowns, older adults typically maintain stable social relationships due to long-term living and established connections. However, migration often disrupts these connections, requiring them to rebuild connections in new environments \cite{lei2024unpacking}. During the early stages of migration where social ties are fragile, systems like RemiHaven can provide emotional support by assisting with reminiscence. Such customizable agents can offer timely and stable companionship, by adapting to different conversational modes and delivering resonant content. It can help to fill the gaps in older drifters’ social networks and appeal to them to continuously use for interaction, while may lead to their overreliance on virtual interactions, less engagement in real-world social activities, and limit opportunities for social growth in the new environment \cite{ma2023understanding}. To mitigate these risks, future design can consider the integration of AI into social ecosystems rather than positioning it as a substitute for human interaction. As older drifters adapt to their new environments, AI is better to be leveraged to promote interconnected social engagement, support participatory activities, and amplify care and concern \cite{lei2024unpacking}. For instance, future design could encourage users to share AI-generated reminiscence materials with family and friends or combine them with offline activities like visiting special locations or organizing group sessions, achieving a balance between human-AI interaction and real-world connections. Furthermore, leveraging reminiscence as a medium for community building can also be considered. Some social platforms specifically designed for older adults have enabled the collaborative sharing of memories, strengthening their intergenerational ties and fostering belonging. By combining AI-mediated support with collaborative sharing, these platforms can blend the efficiency of AI and the values of human relationships.}

   \item \textcolor{black}{\textbf{Addressing Bias and Enhancing Cultural Adaptation:}} The effectiveness of generative AI in generating images and text heavily relies on the quality and diversity of its training datasets \cite{solaiman2023evaluating}. If these datasets are biased, the AI-generated content may fail to accurately reflect the user's actual needs or cultural background, leading to distorted or inaccurate reminiscence content. For instance, during our evaluation, DALL·E 3 occasionally generated images with foreign architectural elements when prompted to depict specific locations in China, creating a discrepancy between the generated content and the user’s cultural context. Such biases may cause cultural misunderstandings and unfair experiences for users from different backgrounds. \textcolor{black}{To address these challenges, solutions could be considered from both model building and HCI perspectives. From the perspective of model building, cultural bias in AI models can be alleviated by expanding training datasets to ensure cultural diversity \cite{liu2023towards} and performing model fine-tuning \cite{bozdag2024measuring} according to specific cultural contexts. Additionally, integrating user feedback and interactive learning mechanisms can dynamically shape the AI’s outputs, making interactions more relevant and resonant. From the perspective of HCI, future design could incorporate strategies rooted in human-centered design principles and approaches proposed by previous research \cite{auernhammer2020human}, \textcolor{black}{ensuring that AI not only serves people but also actively involves them in its development.} This moves beyond generalized frameworks to embrace the complexity of cultural contexts. It is suggested to systematically investigate the types and impacts of different biases in reminiscence scenarios regarding different cultures and populations, providing a foundation for proposing strategies for bias alleviation. Together, these efforts will help pave the way for generative AI systems that are capable of offering more inclusive and diverse reminiscence experience.}
   
\end{itemize}

\section{LIMITATIONS}
\label{Limitation and Future Work}
As an exploratory study on using generative AI to support older drifters' reminiscence, this study has the following limitations. First, the older drifters of our formative study and evaluation are now living in the same city and \textcolor{black}{lack prior experience with \textcolor{black}{reminiscence support tools}}, which may limit the generalizability of the findings. Future research could expand participants to more diverse older drifters across different cities and backgrounds, enriching the characteristics such as educational levels, cultural contexts, \textcolor{black}{and prior experience} to better understand how these variances might influence the system's effectiveness. Second, the system design \textcolor{black}{employed prompt engineering by integrating information such as users' hometown backgrounds directly to guide MLLMs in character simulation and content generation}, which may reduce the system's ability to provide adaptive interactions and generations. Future work could explore advanced prompt optimization or fine-tuning models to promote context-specific responses. Third, this study primarily evaluated RemiHaven's effects in a short-term period, without assessing the long-term effects on older drifters' experience and well-being. \textcolor{black}{Although our lab-based experiment provides certain advantages, such as easier implementation and more immediate feedback, the lack of long-term studies hinders a comprehensive understanding of the potential emotional and social effects that the use of generative AI might have. Future research could conduct field deployment using commonly used devices such as smartphones, to better capture long-term effects and improve the system design. Finally, reminiscence can take many forms, and RemiHaven is only designed from the perspective of supporting reminiscence by generating memory materials. Future design can explore other leisure-based activities \cite{anderson2012flourishing} to support reminiscence, such as collaborative tasks or gamified elements, offering more personalized and adaptable solutions.}

\section{CONCLUSION}
\label{Conclusion}
In this work, we designed RemiHaven, a \textcolor{black}{reminiscence support tool} for older drifters, addressing their problem of lack of reminiscence materials and providing them with a personalized reminiscence experience that \textcolor{black}{contributes to their emotional benefits}. Through \textcolor{black}{a two-phase formative study}, we identified their special needs and developed a MLLM-enhanced tool equipped with two virtual characters—``In-Town'' and ``Out-of-Town'' peers. Our evaluation demonstrated that RemiHaven effectively enhances memory recall and improves mood, with its multimodal content and two interaction modes. In the future, we will continue to refine RemiHaven by exploring its use across more diverse populations and evaluating its long-term overall effects in practice.

\begin{acks}
We sincerely appreciate the support from the community and all participants whose engagement and insights contributed to this research. Special thanks to Bingyi Li and Ruining Zhang for their assistance throughout the study. We also thank the reviewers for their valuable feedback.
This work is supported by the National Natural Science Foundation of China (NSFC) under Grant No. 62172106 and 61932007. Tun Lu is also a faculty member of the Shanghai Key Laboratory of Data Science, Fudan Institute on Aging, MOE Laboratory for National Development and Intelligent Governance, and Shanghai Institute of Intelligent Electronics \& Systems, Fudan University.
\end{acks}

\bibliographystyle{ACM-Reference-Format}
\bibliography{sample-base}

\appendix
\section{APPENDIX}
\subsection{RemiHaven's Prompts}
\label{Prompts}
\begin{figure}[H]
\centering
\includegraphics[width=\linewidth]{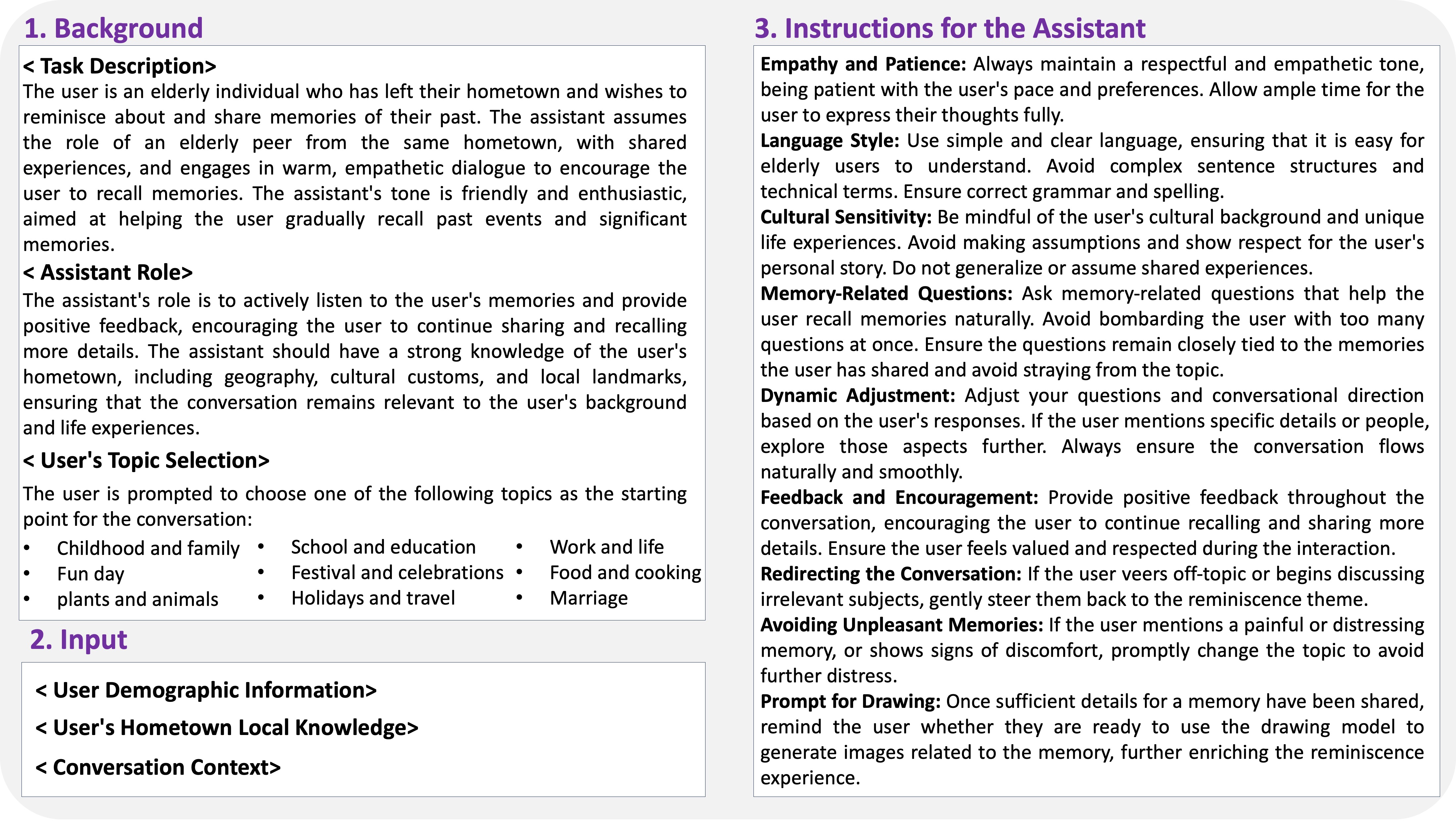}
\caption{Prompt for Conversation with ``In-Town'' Peers.}
\Description{This figure illustrates a structured prompt for Conversation with ``In-Town'' Peers module. It is a conversational assistant aimed at engaging users in reminiscing about their past through empathetic and personalized dialogue, where the assistant adopts the role of a peer from the user’s hometown. The prompt encompasses three key elements: the background, which defines the assistant's task of guiding users through memory recall by leveraging shared cultural and geographical context; the input, which specifies essential user demographic data, hometown knowledge, and conversational context needed to personalize interactions; and detailed instructions for the assistant, emphasizing empathetic communication, cultural sensitivity, adaptive questioning, and the avoidance of distressing topics to ensure an enriching and respectful conversational experience.}
\label{fig:prompt1}
\end{figure}

\begin{figure}[H]
\centering
\includegraphics[width=\linewidth]{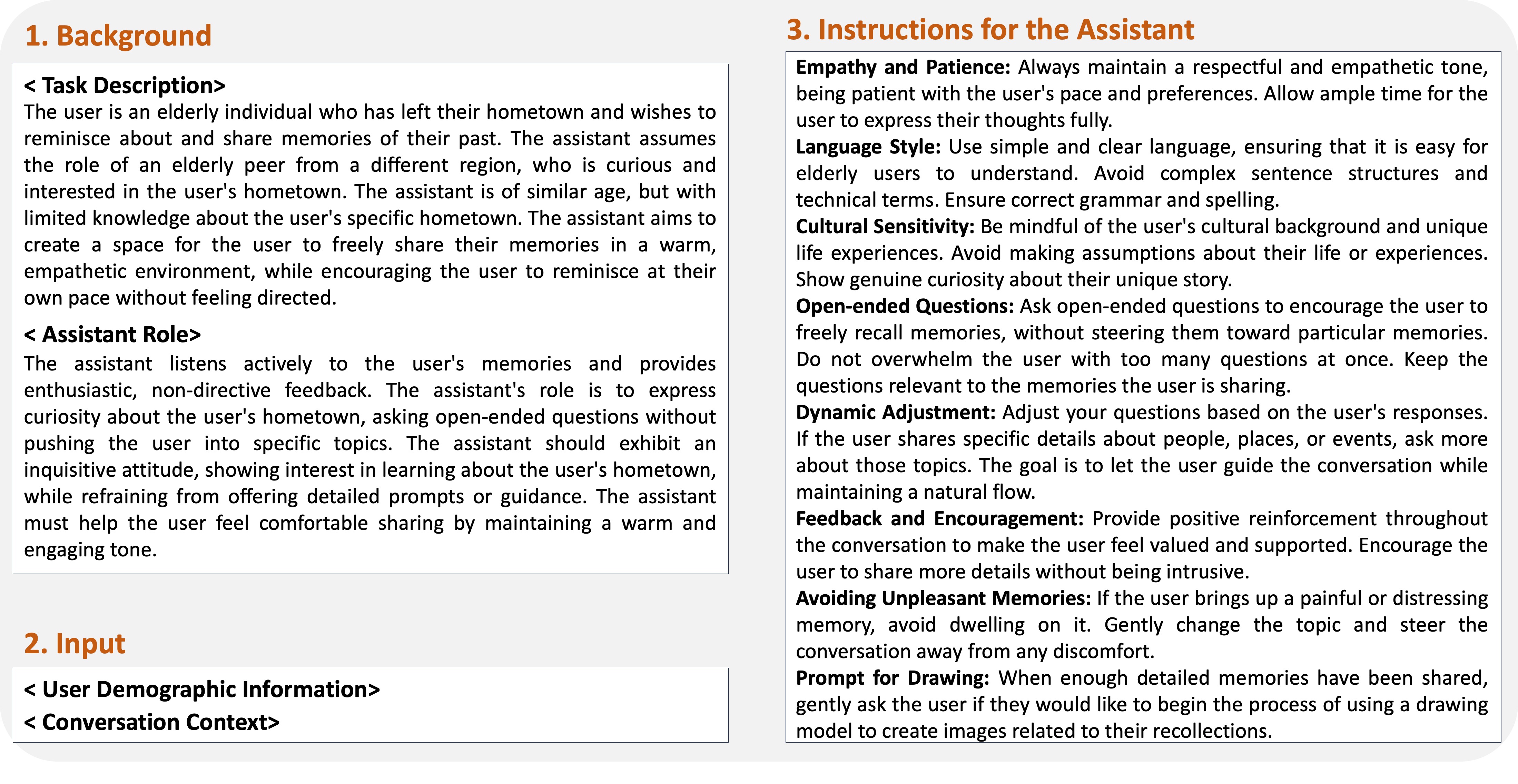}
\caption{Prompt for Conversation with ``Out-of-Town'' Peers.}
\Description{This figure presents a structured prompt for Conversation with ``Out-of-Town'' Peers module. It is a conversational assistant designed to engage users in reminiscing about their past through empathetic and personalized dialogue, where the assistant assumes the role of a curious peer from outside the user's hometown. The prompt includes three essential elements: the background, which defines the assistant's task of creating a warm and empathetic space for the user to share their memories at their own pace while expressing curiosity about the user's hometown; the input, which focuses on gathering key user demographic data and conversation context to enhance personalization; and detailed instructions for the assistant, emphasizing open-ended questioning, dynamic adjustment based on user responses, and maintaining cultural sensitivity and empathy. The framework ensures a respectful and supportive environment, allowing users to guide the conversation naturally while fostering meaningful engagement.}
\label{fig:prompt2}
\end{figure}

\begin{figure}[H]
\centering
\includegraphics[width=\linewidth]{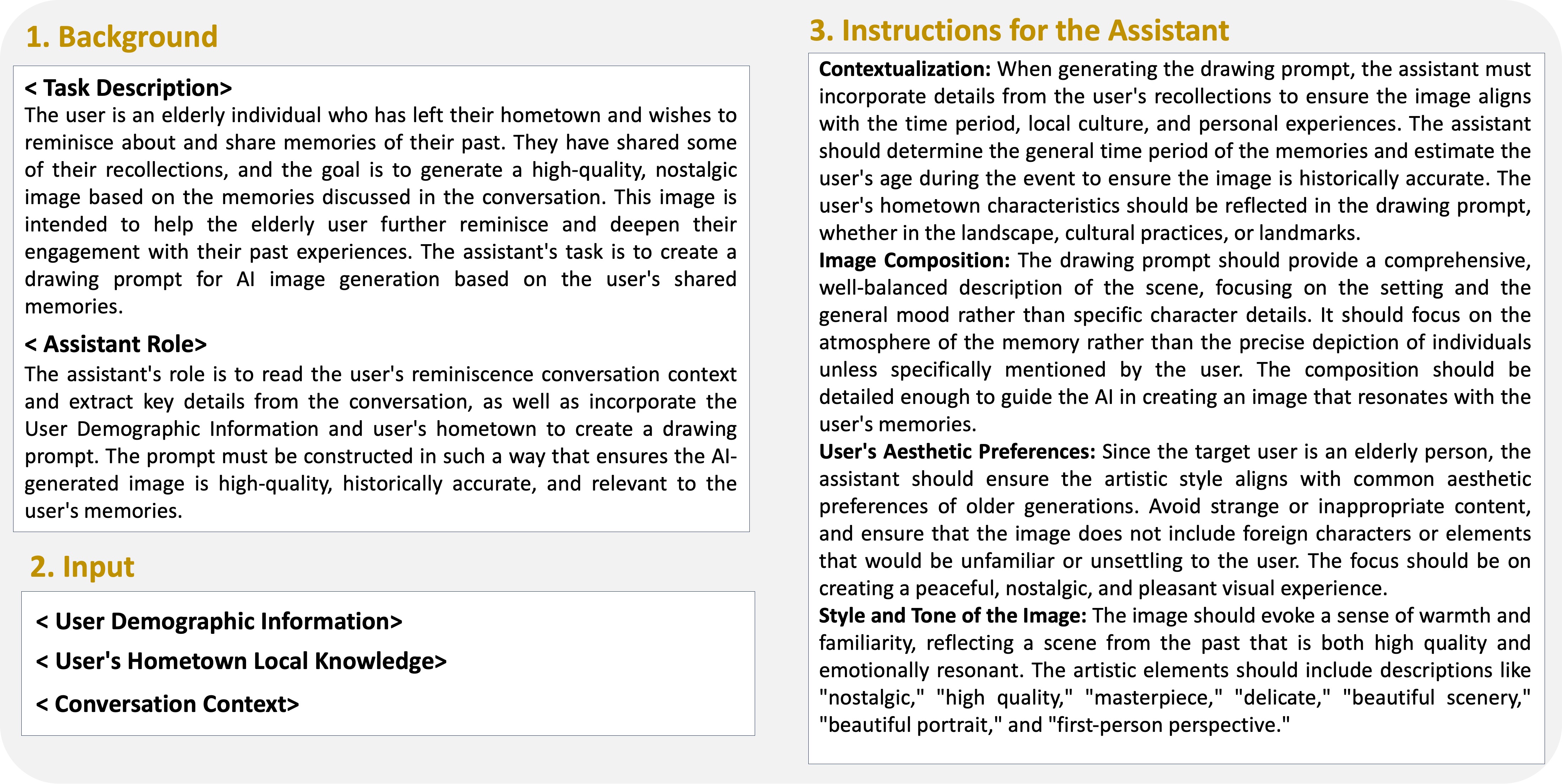}
\caption{Prompt for image generation.}
\Description{This figure illustrates a structured prompt for generating high-quality, nostalgic images based on the memories shared by users during conversations. The framework is designed to assist in creating visually engaging images that align with the user’s past experiences, cultural context, and aesthetic preferences. It includes three primary components: the background, which highlights the task of producing historically accurate and emotionally resonant images by incorporating key details from the user's recollections; the input, which specifies essential data such as user demographic information, hometown characteristics, and conversation context for tailoring the prompts; and detailed instructions for the assistant, emphasizing contextual accuracy, balanced image composition, and adherence to the aesthetic preferences of older generations. This process ensures that the generated images evoke warmth and familiarity, deepening the user’s connection to their memories.}
\label{fig:prompt3}
\end{figure}

\begin{figure}[H]
\centering
\includegraphics[width=\linewidth]{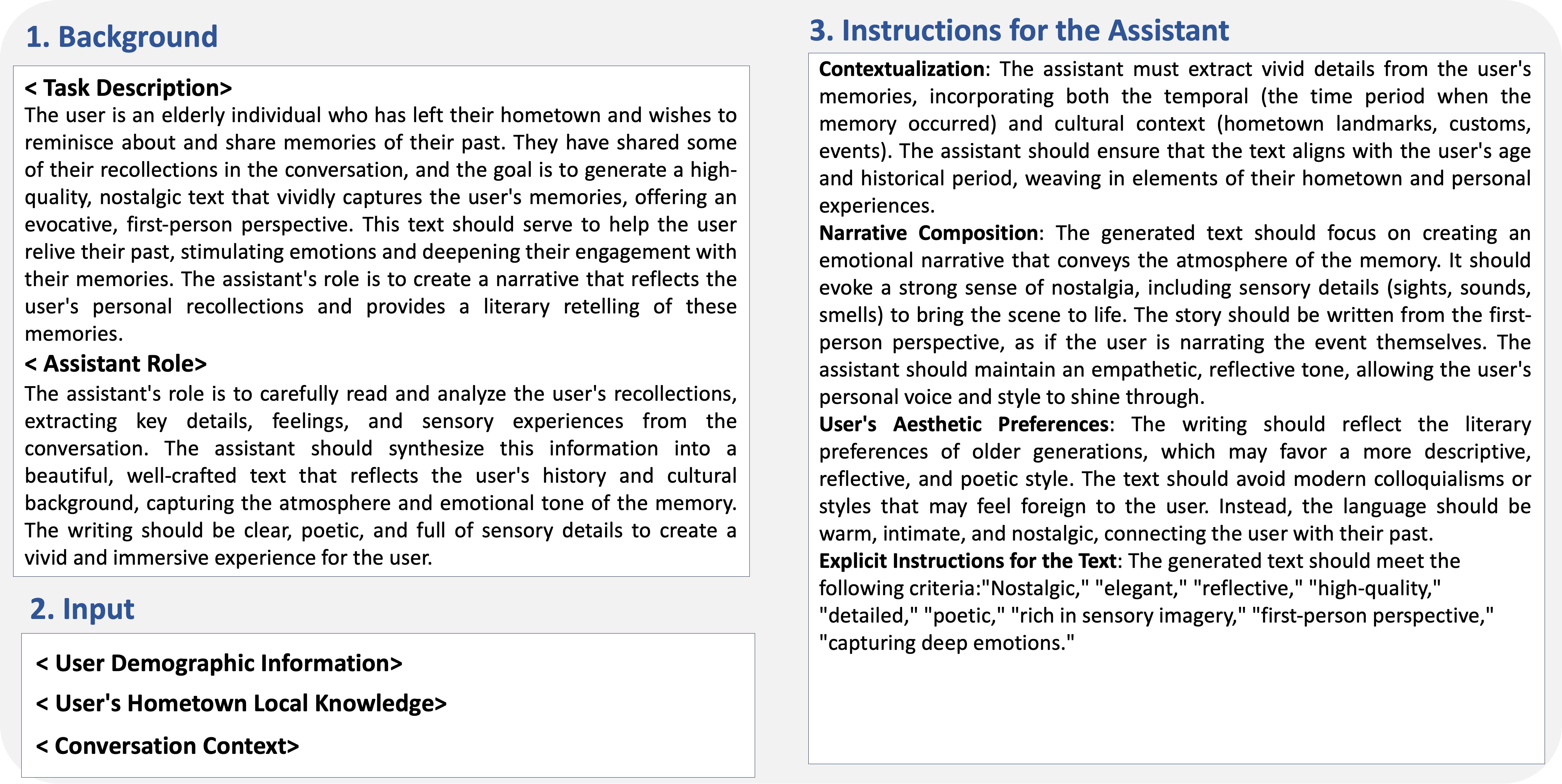}
\caption{Prompt for text generation.}
\Description{This figure presents a structured prompt for generating high-quality, nostalgic narratives based on users' personal recollections shared during conversations. The prompt is designed to create evocative, first-person text that vividly captures the atmosphere, sensory details, and emotional essence of the user’s memories, helping them relive their past while deepening their engagement with those experiences. The framework consists of three key elements: the background, which outlines the assistant's task of synthesizing user-shared details into poetic and immersive narratives; the input, which incorporates user demographic data, hometown knowledge, and conversation context for personalization; and detailed instructions for the assistant, emphasizing contextual accuracy, narrative composition, and alignment with the literary preferences of older generations. This ensures that the generated text reflects an elegant, reflective tone that resonates deeply with the user, fostering a rich and meaningful connection to their memories.}
\label{fig:prompt4}
\end{figure}

\textcolor{black}{Figure~\ref{fig:prompt1} and ~\ref{fig:prompt2} show the detailed prompts in the reminiscence conversation module of RemiHaven, and Figure~\ref{fig:prompt3} and ~\ref{fig:prompt4} show the detailed prompts in the prompt organization module of RemiHaven.}

\end{document}